\documentclass[print,pre,preprintnumbers,twocolumn]{revtex4-2}
\usepackage{amssymb}
\usepackage{dcolumn}
\usepackage{amsmath}
\usepackage{bm}
\usepackage{graphicx}
\usepackage{ulem}
\usepackage[utf8]{inputenc}
\usepackage{amsmath}
\usepackage{amsthm}
\usepackage{bm}

\begin{document}

\title{Quantum thermodynamics with strong system-bath coupling: A mapping
approach}
\author{You-Yang Xu$^{1}$}
\author{Jiangbin Gong$^{2,3,4,5}$}
\email{phygj@nus.edu.sg }
\author{Wu-Ming Liu$^{6}$}
\affiliation{$^{1}$ Faculty of Science, Kunming University of Science and Technology,
Kunming 650500, China\\
$^{2}$ Department of Physics, National University of Singapore, Singapore
117551, Singapore\\
$^{3}$ Center for Quantum Technologies, National University of Singapore,
Singapore 117543, Singapore\\
$^{4}$ Joint School of National University of Singapore and Tianjin
University, International Campus of Tianjin University, Binhai New City,
Fuzhou 350207, China\\
$^{5}$MajuLab, CNRS-UCA-SU-NUS-NTU International Joint Research Unit,
Singapore \\
$^{6}$ Beijing National Laboratory for Condensed Matter Physics, Institute
of Physics, Chinese Academy of Sciences, Beijing 100190, China}

\begin{abstract}
Quantum thermodynamic quantities, normally formulated with the assumption of
weak system-bath coupling (SBC), can often be contested in physical
circumstances with strong SBC. This work presents an alternative treatment
that enables us to use standard concepts based on weak SBC to tackle with
quantum thermodynamics with strong SBC. Specifically, via a
physics-motivated mapping between strong and weak SBC, we show that it is
possible to identify thermodynamic quantities with arbitrary SBC, including
work and heat that shed light on the first law of thermodynamics with strong
SBC. Quantum fluctuation theorems, such as the Tasaki-Crooks relation and
the Jarzynski equality are also shown to be extendable to strong SBC cases.
Our theoretical results are further illustrated with a working example.
\end{abstract}

\maketitle

\textit{Introduction.}---There has been growing fundamental interest in
understanding thermodynamics with strong system-bath coupling (SBC).
In classical thermodynamics, the concept of potential of mean force \cite%
{Kirkwood} was used to facilitate coarse graining to account for strong SBC
\cite{Strasberg}. Corrections to basic classical thermodynamic quantities
due to strong SBC were also known \cite{Seifert}. The classical Jarzynski
fluctuation theorem can be generalized to strong SBC cases \cite{Jarzynski}.
Parallel problems in the quantum domain are more intriguing and can even
lead to divided views. Nevertheless, it is worthwhile highlighting the
following three progresses in quantum thermodynamics treating strong SBC:
(i) the system plus the bath can be thermalized to a Gibbs equilibrium state
of the composite system \cite{Rigol}, (ii) a quantum version of the
potential of mean force may be introduced \cite{squantum,Philipp}; and (iii)
the quantum Tasaki-Crooks fluctuation theorem \cite{Peter} is extendable to
cases with arbitrary SBC strength. There are also other discussions on
possible corrections to quantum thermodynamic quantities due to strong SBC
\cite{Llobet}. However, great caution is always needed when extending
quantum thermodynamics concepts from weak to strong SBC, insofar as a strong
SBC blurs the distinction between system and bath \cite%
{negative,Rivas,Colla,Ivander}.

Notwithstanding possible arbitrariness in defining thermodynamic quantities
in the quantum domain, cases with strong SBC emerge in a variety of
platforms, ranging from condensate systems \cite{Weiss}, quantum optics \cite%
{Lambropoulos}, to quantum biology \cite{Lambert} etc. 
We are hence motivated to seek an innovative quantum thermodynamics approach
that can draw a physical mapping between strong and weak SBC. Indeed, if
there is such a connection between strong and weak SBC, then quantum
thermodynamics with strong SBC can be formulated and digested with standard
concepts based solely on weak SBC. There are two key elements in our
approach. First, the time propagator of the system-bath composite system
giving rise to intricate system-bath dynamics may be transformed to a
product of propagators associated with three stages, including a middle
stage without system-bath interaction. This stage in the absence of SBC can
be exploited to extract work done to the system without any ambiguity.
Secondly, in other two stages, the effect of strong SBC is identified to be
physically equivalent to that of weak SBC, but over an extremely long period
of thermal relaxation. A physics based mapping between weak and strong SBC
is thus achieved. As a result the heat exchange during these two stages can
be clearly identified. 
Consequently, with least assumptions we are able to derive basic
thermodynamic quantities under strong SBC. As compelling evidence of the
usefulness of our mapping approach, at least conceptually, we further show
that under strong SBC, the first law of thermodynamics as a basic
thermodynamic relation can assume a form parallel to that for weak SBC; and
that quantum fluctuation theorems can be extended to cases with strong SBC.
Finally, we also apply our theoretical results to a simple situation to
illustrate the relevance of our approach. 

\begin{figure}[tbph]
\centering\includegraphics[width=3.0in]{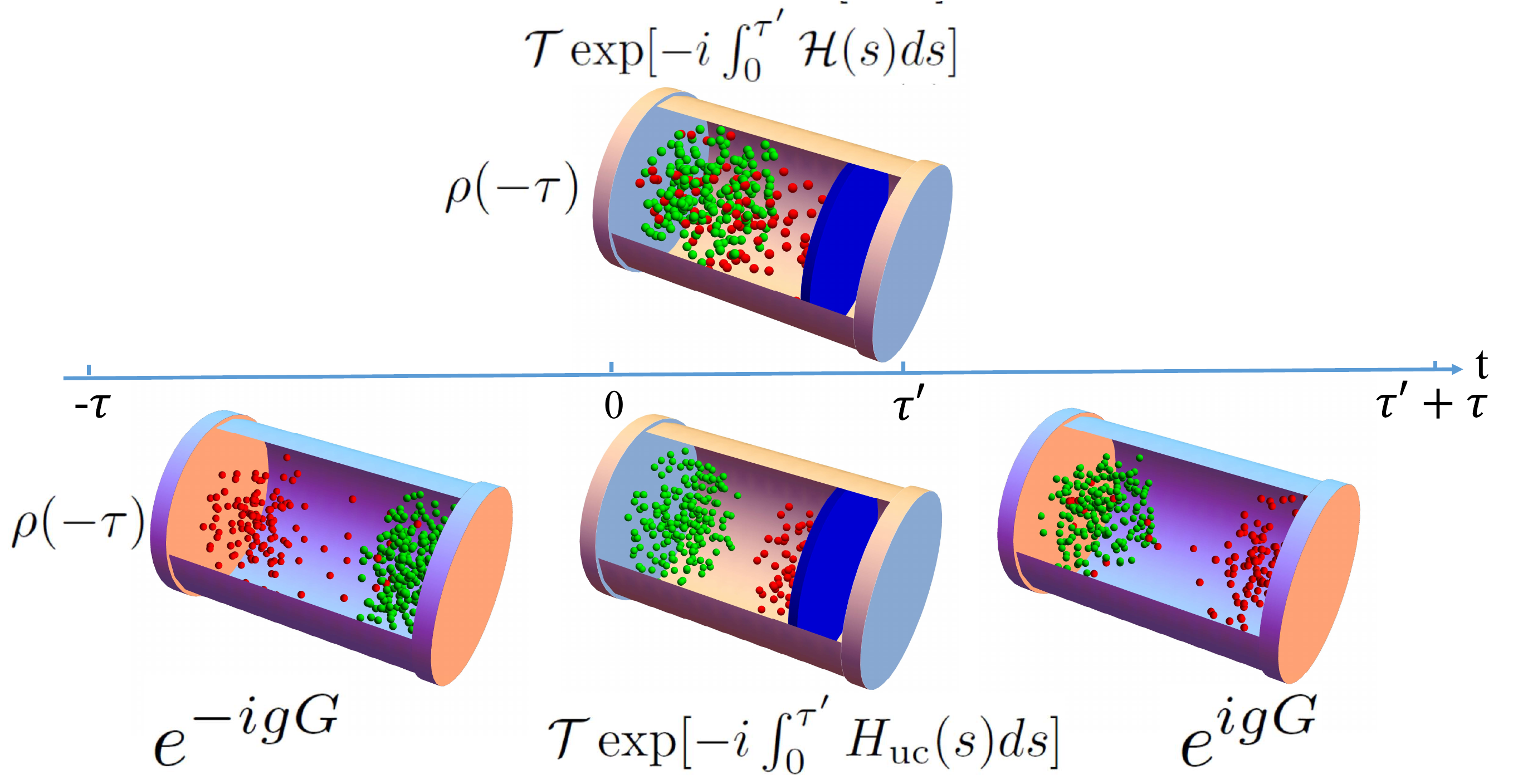}
\caption{Schematic of a three-stage time evolution picture. The time
evolution for an open system with strong SBC during the time interval $[0,%
\protect\tau ^{\prime }]$ is interpreted as the time evolution over $[-%
\protect\tau ,\protect\tau ^{\prime }+\protect\tau ]$ with the same initial
state $\protect\rho (-\protect\tau )$. In this picture, the first and third
stages over the time intervals $[-\protect\tau, 0]$ and $[\protect\tau%
^{\prime }, \protect\tau+\protect\tau^{\prime }]$ involve heat exchange with
a bath under weak SBC without work and the middle stage over the time
interval $[0, \protect\tau^{\prime }]$ involves work without heat. Upon our
mapping detailed in the main text, both Hamiltonians of the system (red
particles) and the bath (green particles) will be corrected by SBC; and the
blue piston represents a work protocol.}
\end{figure}

\textit{Mapping between strong and weak SBC}---Consider the following
composite system-bath Hamiltonian:%
\begin{equation}
\mathcal{H}(t)=\mathcal{H}_{\mathrm{s}}(t)+\mathcal{H}_{\mathrm{I}}(t)+%
\mathcal{H}_{\mathrm{b}}(t),  \label{Eq1}
\end{equation}%
where $\mathcal{H}_{\mathrm{s}}(t)$ and $\mathcal{H}_{\mathrm{b}}(t)$ are
the respective Hamiltonians of the system and the bath, and $\mathcal{H}_{%
\mathrm{I}}(t)=g\sum\limits_{\mathrm{j}}A_{\mathrm{s}}^{\mathrm{j}}(t)A_{%
\mathrm{b}}^{\mathrm{j}}(t)$ depicts the SBC, with system operators $A_{%
\mathrm{s}}^{\mathrm{j}}$ and bath operators $A_{\mathrm{b}}^{\mathrm{j}}$.
Here $g$ is the dimensionless SBC strength, with $g\ll 1$ indicating weak
coupling and $g\gtrsim 1$ indicating strong coupling.

In treating light-matter interaction \cite{Ashida, Gong-Lee} or polaronic
systems \cite{Lee}, a strong system-bath coupling can be connected with
cases of a weak coupling via some unitary transformation. Mimicking but
different from this idea, let us now introduce an unitary transformation $%
e^{igG}$, through which we can connect the original composite system
Hamiltonian $\mathcal{H}(t)$ with a system-bath uncoupled Hamiltonian $H_{%
\mathrm{uc}}$, namely,
\begin{equation}
\mathcal{H}(t)=e^{igG}H_{\mathrm{uc}}(t)e^{-igG},  \label{teq}
\end{equation}%
where the transformation generator $G=\sum\limits_{\mathrm{j}}B_{\mathrm{s}%
}^{\mathrm{j}}B_{\mathrm{b}}^{\mathrm{j}}$ with $B_{\mathrm{s}}^{\mathrm{j}}$
being system operators and $B_{\mathrm{b}}^{\mathrm{j}}$ being bath
operators. The uncoupled Hamiltonian assumes the form of $H_{\mathrm{uc}%
}(t)=H_{\mathrm{s}}(t)+H_{\mathrm{b}}$, with the transformed system
Hamiltonian $H_{\mathrm{s}}(t)$ being time dependent in general, and the
transformed bath Hamiltonian $H_{\mathrm{b}}$ being time independent.

For a given $\mathcal{H}(t)$, it appears to be a challenge to transform, via
the generator $G$ introduced above, to a system-bath uncoupled Hamiltonian $%
H_{\mathrm{uc}}(t)$. Since we are more concerned with fundamentals of
thermodynamic quantities with strong SBC, we allow ourselves to start with $%
H_{\mathrm{uc}}$ instead. Nevertheless, because $G$ is arbitrary, in
principle the transformation $e^{igG}$ suffices to yield a rather generic $%
\mathcal{H}_{\mathrm{I}}$ from $H_{\mathrm{uc}}$. Consider then the
following explicit example where $H_{\mathrm{uc}}=H_{\mathrm{s}}(t)+H_{%
\mathrm{b}}$ with $H_{\mathrm{s}}(t)=\lambda _{\mathrm{x}}(t)\sigma _{%
\mathrm{s}}^{\mathrm{x}}+\lambda _{\mathrm{z}}(t)\sigma _{\mathrm{s}}^{%
\mathrm{z}}$, $\{\sigma ^{i}\}$ the Pauli operators, $\lambda _{\mathrm{x}%
}(t)$ and $\lambda _{\mathrm{z}}(t)$ two time-dependent paramaters, and $H_{%
\mathrm{b}}=-\omega _{\mathrm{b}}\sum_{\mathrm{k}}\sigma _{\mathrm{k}}^{%
\mathrm{z}}-h\sum_{\left\langle \mathrm{k,j}\right\rangle }\sigma _{\mathrm{k%
}}^{\mathrm{x}}\sigma _{\mathrm{j}}^{\mathrm{x}}$, depicting a spin bath
modeled by an Ising model with $N$ spins interacting with the
nearest-neighbor sites $\left\langle \mathrm{k,j}\right\rangle $. Planck's
constant is taken as unity throughout. This uncoupled system can be
connected with a model where a central spin interacts with an Ising bath
\cite{Caldeira,Vega} by considering $G=\sigma _{\mathrm{s}}^{\mathrm{x}%
}\sum_{\mathrm{k}}\sigma _{\mathrm{k}}^{\mathrm{x}}$. Using the
transformation depicted by Eq.~(\ref{teq}) above, one finds $\mathcal{H}_{%
\mathrm{s}}=H_{\mathrm{s}}(t)=\lambda _{\mathrm{x}}(t)\sigma _{\mathrm{s}}^{%
\mathrm{x}}+\lambda _{\mathrm{z}}(t)\sigma _{\mathrm{s}}^{\mathrm{z}}$. $%
\mathcal{H}_{\mathrm{b}}$ differs from the mapped $H_{\mathrm{b}}$, namely, $%
\mathcal{H}_{\mathrm{b}}=-\omega _{\mathrm{b}}\cos (g)\sum_{\mathrm{k}%
}\sigma _{\mathrm{k}}^{\mathrm{z}}-h\sum_{\left\langle \mathrm{kj}%
\right\rangle }\sigma _{\mathrm{k}}^{\mathrm{x}}\sigma _{\mathrm{j}}^{%
\mathrm{x}}$ \cite{surp}. Finally, the original system-bath coupling term $%
\mathcal{H}_{\mathrm{I}}=-\omega _{\mathrm{b}}\sin (g)\sigma _{\mathrm{s}}^{%
\mathrm{x}}\sum_{\mathrm{k}}\sigma _{\mathrm{k}}^{\mathrm{y}}+\lambda _{%
\mathrm{z}}(t)\sigma _{\mathrm{s}}^{\mathrm{z}}[\cos (g\sum_{\mathrm{k}%
}\sigma _{\mathrm{k}}^{\mathrm{x}})-1]+\lambda _{\mathrm{z}}(t)\sigma _{%
\mathrm{s}}^{\mathrm{y}}\sin (g\sum_{\mathrm{k}}\sigma _{\mathrm{k}}^{%
\mathrm{x}})$ \cite{surp}. This working example here has thus illustrated
that it is feasible to connect an original open quantum system Hamiltonian $%
\mathcal{H}(t)$ with a system-bath decoupled Hamiltinian $H_{\mathrm{uc}}(t)$%
. A second example involving a Caldeira-Leggett like model is discussed in
Supplementary Materials \cite{surp}.

The unitary propagator of the original composite system over the time
duration $[0,\tau ^{\prime }]$ can be formally written as $U_{0\rightarrow
\tau ^{\prime }}=\mathcal{T}\exp [-i\int_{0}^{\tau ^{\prime }}\mathcal{H}%
(s)ds]$. Thanks to the transformation introduced in Eq.~(\ref{teq}), $%
U_{0\rightarrow \tau ^{\prime }}$ can be written as a product of three
unitaries: $U_{0\rightarrow \tau ^{\prime }}=e^{igG}U_{\mathrm{uc}}e^{-igG}$%
. In other words, whatever is captured by $U_{0\rightarrow \tau ^{\prime }}$
can be digested in terms of three stages, a unitary process $e^{-igG}$ due
to our transformation, a unitary propagator $U_{\mathrm{uc}}=\mathcal{T}\exp
[-i\int_{0}^{\tau ^{\prime }}H_{\mathrm{uc}}(s)ds]$ with system and bath
uncoupled, and a third unitary $e^{igG}$ \cite{surp} also due to our
transformation. This is schematically shown in Fig. 1. A few important
remarks are in order. First, by construction, $U_{\mathrm{uc}}$ represents
the propagator under $H_{\mathrm{uc}}(t)$. Energy change caused by $U_{%
\mathrm{uc}}$ is entirely due to work done to the system because there is no
system-bath interaction during this step. Secondly, the two unitaries $%
e^{\pm igG}$ are apparently responsible for energy exchange between system
and bath, and hence heat can be extracted from these two processes without
arbitrariness while no work protocol is executed (since there are no changes
to all the system parameters). Specifically, as shown in Supplementary
Materials \cite{surp}, the two unitary operators $e^{-igG}$ and $e^{igG}$
can be interpreted as the respective outcomes of the time evolution of the
composite systems $H_{\mathrm{uc}}(0)$ and $H_{\mathrm{uc}}(\tau ^{\prime })$
over a duration defined as $\tau $, evolving under the time-dependent SBC $%
H_{-}=\frac{g}{\tau }G_{-}$ and $H_{+}=-\frac{g}{\tau }G_{+}$ with $%
G_{-}=\exp [-iH_{\mathrm{uc}}(0)t]G\exp [iH_{\mathrm{uc}}(0)t]$ and $%
G_{+}=\exp [-iH_{\mathrm{uc}}(\tau ^{\prime })t]G\exp [iH_{\mathrm{uc}}(\tau
^{\prime })t]$ \cite{surp}. If we now choose $\tau $ to be sufficiently
large, then $e^{-igG}$ and $e^{igG}$ represents a slow relaxation process
experienced by some quantum open system with weak SBC. This way, time is
exploited to uncover a physics-based mapping between strong and weak SBC. Of
particular interest is to consider extremely long processes ($\tau
\rightarrow +\infty $), such that on the one hand, we can safetly apply our
knowledge in the weak SBC regime and on the other hand, one can assume
self-relaxation of the composite system $H_{\mathrm{uc}}$ towards its Gibbs
state under the actions of $e^{-igG}$ and $e^{igG}$. One may be concerned
that turning on the $H{\pm }$ introduced in our picture above can involve
work as well. This concern can be lifted because the associated work due to $%
H_{\pm }$ is inversely proportional to $\tau $ \cite{surp}. In the limit of
very large $\tau $, such kind of work due to the switching on of SBC
vanishes.

\textit{Quantum thermodynamics at arbitrary SBC strength}---With the above
three-stage picture (see Fig.~1), it is now possible to identify basic
thermodynamic quantities at arbitrary SBC strength, by use of
well-established quantities for weak SBC \cite{Alicki,Talkner,Campisi,M
Esposito,two,dtcf}. As stated above, heat exchange can be calculated based
on the propagator associated with the two time intervals $[-\tau ,0]$ and $%
[\tau ^{\prime },\tau ^{\prime }+\tau ]$. There is no conceptual difficulty
in doing so because the composite system Hamiltonian is understood to be a
static one, namely $H_{\mathrm{uc}}(0)$ or $H_{\mathrm{uc}}(\tau^{\prime })$
during these two stages and hence no work is involved by construction. To
arrive at a differential form of heat, we can also take duation $\tau
^{\prime }$ to be infinitesimal. We then find that, for an arbitrary
coupling strength $g$, heat exchange is given by the following expression
\cite{surp}:
\begin{equation}
dQ=\mathrm{Tr}_{\mathrm{s}}(H_{\mathrm{s}}~d\rho _{\mathrm{s}})+\mathrm{Tr}%
[\rho ~d(H_{\mathrm{s}}-\mathcal{H})],  \label{dqeq}
\end{equation}%
where $\mathrm{Tr}_{\mathrm{s}}(A)$ and $\mathrm{Tr(}A)$ are trace
operations over the system and the sytem-bath composite system,
respectively. $\rho $ here represents the true initial state of the
composite system and $\rho _{\mathrm{s}}$ is the reduced density of the
system. The second term in the equation above represents a correction due to
SBC, which, as expected, vanishes in cases of vanishingly weak SBC.

Likewise, to identify work done to the system we now turn our attention to
the time interval $[0,\tau ^{\prime }]$. According to the above three-stage
picture again, during the time interval $[0,\tau ^{\prime }]$ the system is
uncoupled with the bath. Thus, the associated heat transfer is necessarily
zero. The work at arbitrary SBC strength is then found to be \cite%
{surp,quenching}:
\begin{equation}
dW=\mathrm{Tr}(\rho~ d\mathcal{H}).  \label{dweq}
\end{equation}%
This expression is obtained from a standard calculation of the expectation
value of quantum work for an isolated system uncoupled from a bath, fully
consistent with the two-time measurement definition of quantum work if the
initial state of the work protocol during the time interval $[0,\tau
^{\prime }]$ possesses no energy coherence \cite{Gong15}.

It is now curious to investigate how the first law of thermodynamics is
manifested in cases with strong SBC. Using Eqs.~(\ref{dqeq}) and (\ref{dweq}%
), one finds
\begin{equation}
dE=dW+dQ=d[\mathrm{Tr}_{\mathrm{s}}(H_{\mathrm{s}}\rho _{\mathrm{s}})]
\label{1stlaw}
\end{equation}%
for arbitrary SBC strength. This result is by no means trivial because $H_{%
\mathrm{s}}$ in Eq.~(\ref{1stlaw}) represents the system component of $H_{%
\mathrm{uc}}$, which is yet to be worked out in order to do explicit
calculations. Because one can still imagine a ficticious $H_{\mathrm{s}}$
before it can be found, the first law of thermodynamics under strong SBC is
physically intriguing.

Having discussed the first law, next we examine the entropy function $S$ in
the presence of strong SBC. To facilitate our investigations let us now
specifically consider a quasi-static process such as the composite system
stays in its Gibbs state, $\Omega (t)=\exp [-\beta \mathcal{H}(t)]/Z(t)$,
where $Z(t)$ is the corresponding partition function and $\beta $ the fixed
inverse temperature of a super bath. Likewise, refering to $H_{\mathrm{uc}%
}=H_{\mathrm{s}}(t)+H_{\mathrm{b}}$, we use $Z_{\mathrm{s}}(t)$ to represent
the partition function associated with $H_{\mathrm{s}}(t)$ and $Z_{\mathrm{b}%
}$ the bath partition function associated with the time-independent bath
Hamiltonian $H_{\mathrm{b}}$. The initial state of the composite system is
then $\Omega (0)=\exp [-\beta \mathcal{H}(0)]/Z(0).$ We also define $\varpi
(t)\equiv \exp [-\beta H_{\mathrm{uc}}(t)]/Z_{\mathrm{uc}}(t)$, with the
associated partition function $Z_{\mathrm{uc}}(t)$ for the uncoupled
Hamiltonian $H_{\mathrm{uc}}$. Using the expression Eq.~(\ref{dqeq}) for $dQ$%
, we obtain
\begin{equation}
dS=\beta dQ=-d[\mathrm{Tr}(\Omega _{\mathrm{s}}\ln \varpi _{\mathrm{s}%
})]-d\ln \frac{Z_{\mathrm{s}}}{Z},  \label{seq}
\end{equation}%
where $\Omega _{\mathrm{s}}=\mathrm{Tr}_{\mathrm{b}}(\Omega )$ and $\varpi _{%
\mathrm{s}}=\mathrm{Tr}_{\mathrm{b}}(\varpi )$ are the reduced state of the
system after tracing $\Omega $ and $\varpi $ over the bath, respectively.
Further using $Z(t)=\mathrm{Tr}[e^{-\beta \mathcal{H}(t)}]=\mathrm{Tr}%
[e^{igG}e^{-\beta H_{\mathrm{uc}}(t)}e^{-igG}]=\mathrm{Tr}[e^{-\beta H_{%
\mathrm{uc}}(t)}]=Z_{\mathrm{uc}}(t)=Z_{\mathrm{s}}(t)Z_{\mathrm{b}}$, the
relation $\frac{Z(t)}{Z_{\mathrm{s}}(t)}=Z_{\mathrm{b}}$ becomes evident and
so the term $d\ln \frac{Z_{\mathrm{s}}}{Z}$ in Eq.~(\ref{seq}) vanishes,
yielding
\begin{equation}
dS=-d\mathrm{Tr}(\Omega _{\mathrm{s}}\ln \varpi _{\mathrm{s}}).
\end{equation}%
This expression for $dS$ can be further rewritten as
\begin{equation}
dS=dS_{v}(\Omega _{\mathrm{s}})+dD(\left. \Omega _{\mathrm{s}}\right\vert
\varpi _{\mathrm{s}}),
\end{equation}%
where the von Neumann entropy $S_{v}(\Omega _{\mathrm{s}})=-\mathrm{Tr}_{%
\mathrm{s}}(\Omega _{\mathrm{s}}\ln \Omega _{\mathrm{s}})$, and the relative
entropy is $D(\left. \Omega _{\mathrm{s}}\right\vert \varpi _{\mathrm{s}})=%
\mathrm{Tr}_{\mathrm{s}}(\Omega _{\mathrm{s}}\ln \Omega _{\mathrm{s}})-%
\mathrm{Tr}_{\mathrm{s}}(\Omega _{\mathrm{s}}\ln \varpi _{\mathrm{s}})$. We
shall return to this useful definition later. For general nonequilbrium
situations, the change in the thermal entropy cannot be calculated from the
time-evolving non-equilibrium state. Nevertherless, one still wishes to
define a change in some information entropy measure. One possible quantity
to measure information entropy change for nonequilibrium cases with
arbitrary SBC strength can be $\Delta S=\Delta S_{v}(\rho _{\mathrm{s}%
})+\Delta D(\left. \Omega _{\mathrm{s}}\right\vert \varpi _{\mathrm{s}})$,
which differs from the quasi-static situations by replacing the Gibbs state
in a quasi-static process with the actual reduced state of the system $\rho
_{\mathrm{s}}$ in a general process.

\textit{Fluctuation theorems with arbitrary SBC strength}-- Our treatment
outlined above makes it straightforward to investigate the work fluctuation
theorems since the work done can be calculated entirely by the middle step
of the above outlined three-stage time evolution. That is, work can be
calculated during the $[0,\tau ^{\prime }]$ interval only, with the mapped
system $H_{\mathrm{s}}(t)$ entirely decoupled from any bath. As such, work
fluctuations of the orginal problem become that of the transformed system
Hamiltonian $H_{\mathrm{s}}(t)$, of which the time dependence does indicate
a possible work protocol. To proceed with explicit calculations, we assume
that the initial state of the original problem is still the Gibbs state $%
\Omega (0)$. During the period of $[-\tau ,0]$ that could be arbitrarily
long, $\Omega (0)$ evolves to $\varpi (0)$ at the end of the first stage or
at the beginning of the middle stage. The so-called characteristic function
of work can then be calculated from $\Theta _{0\rightarrow \tau ^{\prime
}}(u)=\int dw~e^{iuw}p_{0\rightarrow \tau ^{\prime }}(w)$, where $%
p_{0\rightarrow t}(w)$ is the probability density of work value $w$ in the
middle stage. 
Adopting the standard treatment in the literature \cite{Talkner}, we can
express $\Theta _{0\rightarrow \tau ^{\prime }}(u)$ in terms of the
propagator $U_{\mathrm{uc}}$ associated with $[0,\tau ^{\prime }]$. That is,
$\Theta _{0\rightarrow \tau ^{\prime }}(u)=\mathrm{Tr}\left[ U_{\mathrm{uc}%
}^{\dagger }e^{iuH_{\mathrm{uc}}(\tau ^{\prime })}U_{\mathrm{uc}}e^{-(\beta
+iu)H_{\mathrm{uc}}(0)}\right] /{Z_{\mathrm{uc}}(0)}$. Further using $%
U_{0\rightarrow \tau ^{\prime }}=e^{igG}U_{\mathrm{uc}}e^{-igG}$, $\mathcal{H%
}(t)=e^{igG}H_{\mathrm{uc}}(t)e^{-igG}$, and $Z(t)=Z_{\mathrm{uc}}(t)$, one
directly obtains
\begin{equation}
\Theta _{0\rightarrow \tau ^{\prime }}(u)=\mathrm{Tr}\left[ U_{0\rightarrow
\tau ^{\prime }}^{\dagger }e^{iu\mathcal{{H}(\tau ^{\prime })}%
}U_{0\rightarrow \tau ^{\prime }}e^{-(\beta +iu)\mathcal{H}(0)}\right] /{Z(0)%
}.  \label{fluceq}
\end{equation}%
Equation (\ref{fluceq}) suggests that work fluctuation theorems are
extendable to cases with arbitrary SBC strength. In particular, given that $%
U_{0\rightarrow \tau ^{\prime }}=U_{\tau ^{\prime }\rightarrow 0}^{\dagger }$%
, one immediately derives from Eq.~(\ref{fluceq}) via an inverse Fourier
transformation the following Tasaki-Crooks quantum fluctuation relation:
\begin{equation}
\frac{p_{0\rightarrow \tau ^{\prime }}(w)}{p_{\tau ^{\prime }\rightarrow
0}(-w)}=\frac{Z(\tau ^{\prime })}{Z(0)}e^{\beta w}.  \label{fluceq2}
\end{equation}%
This quantum fluctuation theorem much resembles to that for weak SBC. It
should be highlighted that the partition function in the above expression is
that of the composite total system, an intriguing result echoing with a
previous elegant result \cite{Peter}. Furthermore, as already suggested in
our use of $Z_{\mathrm{uc}}(t)=Z(t)$ in obtaining Eq.~(\ref{fluceq2}) above,
the above Tasaki-Crooks quantum fluctuation can be also expressed as
\begin{equation}
\frac{p_{0\rightarrow \tau ^{\prime }}(w)}{p_{\tau ^{\prime }\rightarrow
0}(-w)}=\frac{Z_{\mathrm{s}}(\tau ^{\prime })}{Z_{\mathrm{s}}(0)}e^{\beta w},
\label{fluceq3}
\end{equation}%
where $Z_{\mathrm{s}}(t)$ is the system partition function of $H_{\mathrm{s}%
}(t)$, the system part of the mapped uncoupled Hamiltonian $H_{\mathrm{uc}%
}(t)$.

One can further extend the celebrated Jarzynski equality to cases with
arbitrary SBC strength. Indeed, one can make use of Eq.~(\ref{fluceq3}) by
integrating $e^{-\beta w}p_{0\rightarrow \tau ^{\prime }}(w)$ over all
possible work values $w$, arriving at
\begin{equation}
\langle e^{-\beta w}\rangle =\frac{Z_{\mathrm{s}}(\tau ^{\prime })}{Z_{%
\mathrm{s}}(0)}\equiv e^{-\Delta F}.
\end{equation}%
In the expression above, we have also defined the free energy $F$ associated
with $H_{\mathrm{s}}$ when it is in thermal equibrium at temperature $T$,
namely, $F\equiv -T\ln Z_{\mathrm{s}}$ and $\Delta F=-T[\ln Z_{\mathrm{s}%
}(\tau ^{\prime })-\ln Z_{\mathrm{s}}(0)]$. Interestingly, with this
definition and our early defined entropy $\Delta S$, one finds $\Delta
F=\Delta E-T\Delta S$ \cite{surp}, which is again analgous to the
corresponding thermodynamic relation for a constant-temperature process
under weak SBC.

\begin{figure}[tbph]
\centering\includegraphics[width=3.4in]{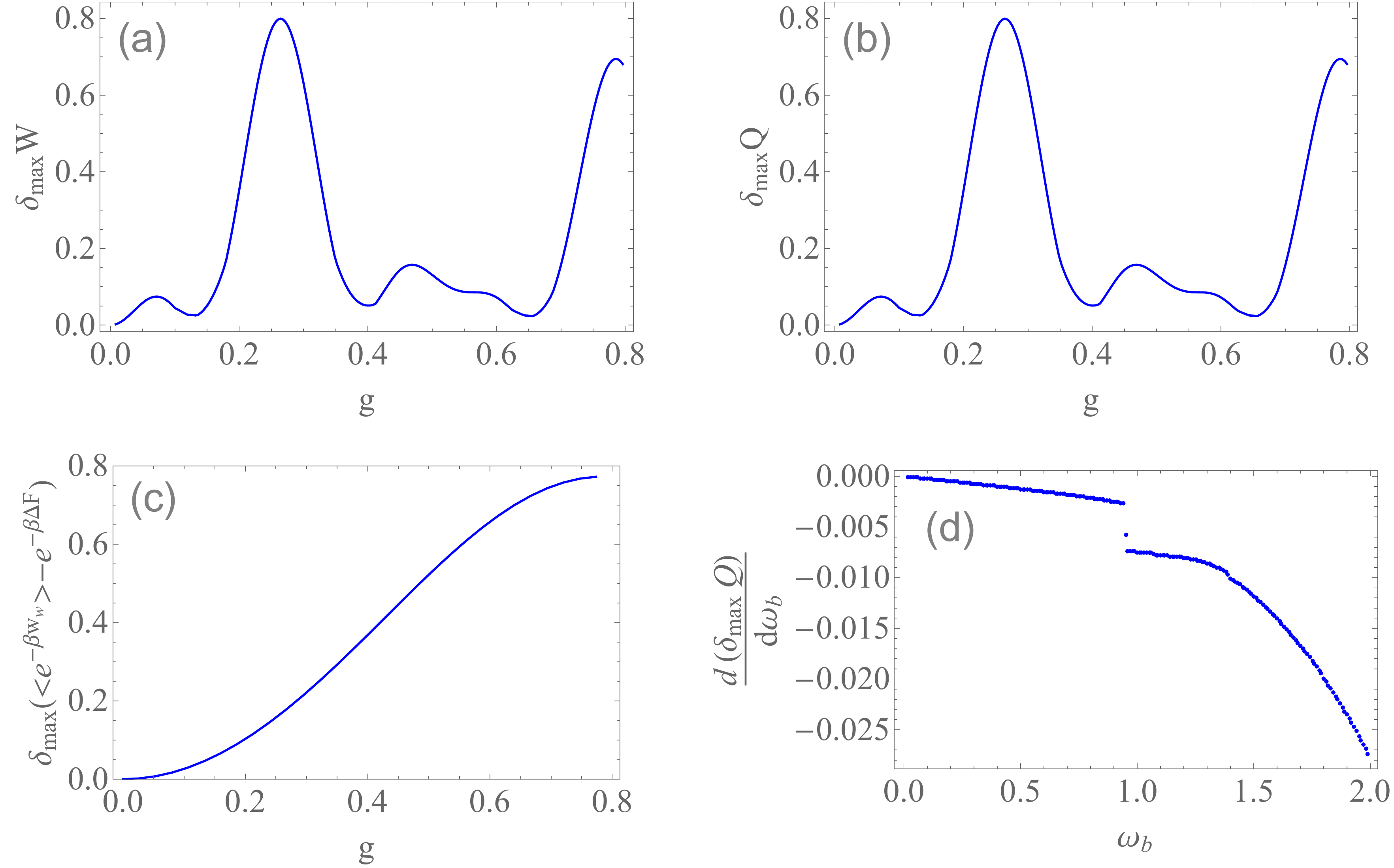}
\caption{Difference between the strongly and weakly coupled quantum
thermodynamics. (a), (b) and (c), respectively, show $\protect\delta _{\max
}W$, $\protect\delta _{\max }Q$ and $\protect\delta _{\max }(\left\langle
e^{-\protect\beta w_{\mathrm{w}}}\right\rangle -e^{-\protect\beta \Delta F})$%
\ as a function of coupling strength $g$ with $\protect\omega _{\mathrm{b}}=1
$; (d) show $\frac{d(\protect\delta _{\max }Q)}{d\protect\omega _{\mathrm{b}}%
}$ as a function of $\protect\omega _{\mathrm{b}}$ with\ coupling strength $%
g=0.1$. The initial state is the Gibbs state $\Omega $ with the invere
temperature $\protect\beta =1$. The other parameters are chosen as, $h=1$, $%
\protect\lambda _{\mathrm{x}}(0)=1$, $\protect\lambda _{\mathrm{z}}(0)=2.5$,
ramping rate $\protect\alpha _{\mathrm{x}}=1$, $\protect\alpha _{\mathrm{z}%
}=-0.6$, number of bath spins $N=6$ and duration time $\protect\tau ^{\prime
}=2$.}
\end{figure}

\textit{Numerical Results}---We further use the central spin model presented
above as an example. This model is much similar to that in \cite{quan} to
study quantum criticality vs the Loschmidt echo if the parameter $\lambda _{%
\mathrm{z}}$ is zero. The system-bath coupling term $\mathcal{H}_{\mathrm{I}%
} $ is equivalent to the Heisenberg XY interaction if the parameter $g$ is
small \cite{xy interaction}. In our simulation, the system is driven through
following protocol, i.e., $\lambda _{\mathrm{z}}(t)=\lambda _{\mathrm{z}%
}(0)(1+\alpha _{z}t)$ and $\lambda _{\mathrm{x}}(t)=\lambda _{\mathrm{x}%
}(0)(1+\alpha _{x}t)$ with the ramping rate $\alpha _{z}$ and $\alpha _{x}$.
The uncoupled bath upon the mapping is a simple one-dimensional Ising model.

We calculate numerically the absolute differences in work and heat between
what we defined with our approach and a direct but incorrect application of
the weak SBC expression, during the time interval $[0,t]\subset \lbrack
0,\tau ^{\prime }]$. We denote such differences by $\delta W $ and $\delta Q$
\cite{diff}. We then record the maximal differences during $[0,\tau ^{\prime
}]$ to measure the effect of SBC, denoted as $\delta _{\max }W$ and $\delta
_{\max }Q$, respectively. Figure 2(a) and 2(b) depicts $\delta _{\max }W$
and $\delta _{\max }Q$ as a function of the SBC strength $g $. It is seen
that as the SBC strength $g$ increases, the basic quantities such as work
and heat are more affected accordingly.

The impact of an increasing SBC strength on work values has a profound
implication for our understanding of the the Jarzynski equality. As a known
result, the second law $\left\langle W\right\rangle \geq \Delta F$ can be
recovered from the Jarzynski equality by the inequality $\left\langle
e^{y}\right\rangle \geq e^{\left\langle y\right\rangle }$, provided that the
involved work values and free energy quantities are physically valid. It is
thus interesting to ask what happens if one naively calculate the weak-SBC
work from $w_{\mathrm{w}}=\int\limits_{0}^{\tau ^{\prime }}\mathrm{Tr}_{%
\mathrm{s}}~(\rho _{\mathrm{s}}\ dH_{\mathrm{s}})$ (which is valid only for
weak SBC) and then compare $\left\langle e^{-\beta w_{\mathrm{w}%
}}\right\rangle $ and $e^{-\beta \Delta F}$ through $\delta _{\max
}(\left\langle e^{-\beta w_{\mathrm{w}}}\right\rangle -e^{-\beta \Delta F})$%
, as shown in Fig.~2(c). Clearly, we see that ensemble average of $e^{-\beta
w_{\mathrm{w}}}$ does not yield a fixed quantity determined by a free energy
difference. If this were true, one might even deduce the possibility of a
violation of the second law. This serious situation is rectified if we
replace the wrongly calculated $w_{\mathrm{w}}$ by the work $w$ we derived
above, resulting in a Jayzynski equality again, independent of the SBC
strength.

It is also intriguing to note that the uncoupled bath $H_{\mathrm{b}}$
accommodate a quantum phase transition at the critical point $h=\omega _{%
\mathrm{b}}$. This phase transition is also detected by the sudden jump in
the rate of change of the quantity $\delta _{\max }Q$ with respect to $%
\omega _{\mathrm{b}}$ (i.e., $d(\delta _{\max }Q)/d\omega _{\mathrm{b}}$),
as shown in Fig.~2(d).

\textit{Conclusion.}---We have theoretically presented an innovative
approach to the treatment of open quantum systems with strong SBC, by
mapping such systems to cases without system-bath coupling at all. Our
approach makes it possible, at least for one wide class of problems where
the said mapping can be assumed, to define and calculate thermodynamic
quantities with arbitrary SBC, including work and heat. The required mapping
might not be easy to find, but even without finding it explicitly, our
approach is of theoretical interest because it indicates the existence of
well-defined thermodynamic quantities and allows for extensions of quantum
fluctuation theorems to arbitrary SBC strength. Our theoretical results are
further illustrated with a working example, where the impact of strong SBC
can be quantitatively studied.

\textit{Acknowledgments.}--Valuable discussions with Mang Feng, Jian-Hui
Wang and Bei-Lok Hu are gratefully acknowledged. Research by J.G. is
supported by the National Research Foundation, Singapore and A*STAR under
its CQT Bridging Grant. This work was also funded by National Key R\&D
Program of China under grants No. 2021YFA1400900, 2021YFA0718300,
2021YFA1400243, NSFC under grants Nos. 11965012, 61835013, 12234012, Space
Application System of China Manned Space Program and Yunan Province's
Hi-tech Talents Recruitment Plan No. YNWR-QNBJ-2019-245.

\pagebreak

\onecolumngrid

\vspace{1cm} 
\begin{large}
\begin{center}{\bf Supplementary Material}\end{center}
\end{large}
\renewcommand{\thefigure}{S\arabic{figure}}
\renewcommand{\theequation}{S\arabic{equation}}
\setcounter{equation}{0}
\setcounter{figure}{0}

\section{Details of the three-stage picture of the time evolution of quantum
open systems with strong system-bath coupling}

Based on Eq.~(2) in the maintext, the time evolution operator of the
original system $\mathcal{H}$ can be written as
\begin{equation}
U_{0\rightarrow \tau ^{\prime }}=e^{igG}U_{\mathrm{uc}}e^{-igG}.
\end{equation}%
Next we can view the unitary operators $e^{-igG}$ and $e^{+igG}$ as two
respective unitary operators associated with Hamiltonians $\pm G$ in some
interaction representation respectively associated with $H_{\mathrm{uc}}(0)$
and $H_{\mathrm{uc}}(\tau ^{\prime })$. With this understanding, we can
proceed to rewrite the above expression into the following product:
\begin{equation}
U_{0\rightarrow \tau ^{\prime }}=U_{\mathrm{w\tau }^{\prime }}U_{\mathrm{uc}%
}U_{\mathrm{w0}},  \label{w0neq}
\end{equation}%
where
\begin{eqnarray}
U_{\mathrm{w0}} &=&\exp [iH_{\mathrm{uc}}(0)\tau ]\mathcal{T}\left[ \exp
\{-i\int_{0}^{\tau }[H_{\mathrm{uc}}(0)+\frac{g}{\tau }G_{-}(s)]ds\}\right]
\label{neq1} \\
U_{\mathrm{w\tau }^{\prime }} &=&\exp [iH_{\mathrm{uc}}(\tau ^{\prime })\tau
]\mathcal{T}\left[ \exp \{-i\int_{0}^{\tau }[H_{\mathrm{uc}}(\tau ^{\prime
})-\frac{g}{\tau }G_{+}(s)]ds\}\right] .  \label{neq2}
\end{eqnarray}%
with
\begin{eqnarray}
G_{-}(t) &=&\exp [-iH_{\mathrm{uc}}(0)t]G\exp [iH_{\mathrm{uc}}(0)t] \\
G_{+}(t) &=&\exp [-iH_{\mathrm{uc}}(\tau ^{\prime })t]G\exp [iH_{\mathrm{uc}%
}(\tau ^{\prime })t].
\end{eqnarray}%
The above relations become more obvious if one considers the following
equalities
\begin{eqnarray}
\tilde{U}_{\mathrm{w0}}(t) &=&\mathcal{T}\left[ \exp \{-i\int_{0}^{t}[H_{%
\mathrm{uc}}(0)+\frac{g}{\tau }G_{-}(s)]ds\}\right] ;  \notag \\
&=&\exp [-iH_{\mathrm{uc}}(0)t]\exp [-i\frac{g}{\tau }Gt] \\
\tilde{U}_{\mathrm{w\tau }^{\prime }}(t) &=&\mathcal{T}\left[ \exp
\{-i\int_{0}^{t}[H_{\mathrm{uc}}(\tau ^{\prime })+\frac{g}{\tau }%
G_{+}(s)]ds\}\right]  \notag \\
&=&\exp [-i[H_{\mathrm{uc}}(\tau ^{\prime })t]\exp [i\frac{g}{\tau }Gt],
\end{eqnarray}%
both of which can be directly proven by differentiating $\tilde{U}_{\mathrm{%
w0}}(t)$ and $\tilde{U}_{\mathrm{w\tau }^{\prime }}(t)$ defined above with
respect to variable $t$.

It is now interesting to comment on the three exponentials in Eq.~(\ref%
{w0neq}). First of all, the evolution operators $\exp [iH_{\mathrm{uc}%
}(0)\tau ]$ and $\exp [iH_{\mathrm{uc}}(\tau ^{\prime })\tau ]$ as
prefactors of $U_{\mathrm{w0}}$ and $U_{\mathrm{w\tau ^{\prime }}}$ shown in
Eq.~(\ref{neq1}) and Eq.~(\ref{neq2}) are referring to evolution operators
of certain fixed system-bath decoupled Hamiltonians: hence neither work or
heat will be involved. Seondly, as the chosen $\tau $ becomes very long, the
time-dependent Hamiltonians $H_{\mathrm{uc}}(0)+\frac{g}{\tau }G_{-}(t)$ or $%
H_{\mathrm{uc}}(\tau ^{\prime })-\frac{g}{\tau }G_{+}(t)$ used in the
expressions for Eq.~(\ref{neq1}) and Eq.~(\ref{neq2}) have vanishingly small
(though time-dependent) system-bath coupling: hence only heat will be
involved in such processes. That is, the work needed to turn on the
time-dependent Hamiltonians are $W_{\pm }=\mp \int \mathrm{Tr}(\frac{g}{\tau
}\rho\ dG_{\pm })=\mp \frac{g}{\tau }\Delta \mathrm{Tr}(\rho G_{\pm })$,
which will be vanishing in the limit $\tau \rightarrow \infty $. This
completes our three-stage time evolution picture: $U_{\mathrm{w\tau }%
^{\prime }}$ and $U_{\mathrm{w0}}$ will incur heat exchange whereas $U_{%
\mathrm{uc}}$ will be responsible for work. This way, we have identified one
clear-cut route towards explicit calculations of heat and work without any
ambiguity.

\section{Models}

To elaborate on the mapping between strong and weak SBC in more detail, we
first consider a central spin-$\frac{1}{2}$ interacting with a spin bath, an
example mainly used by the main text. According to the main text, the
system-bath uncoupled Hamiltonian upon a desired mapping is%
\begin{equation}
H_{\mathrm{uc}}=H_{\mathrm{s}}(t)+H_{\mathrm{b}},  \label{s9}
\end{equation}%
where the central spin Hamiltonian is $H_{\mathrm{s}}(t)=\lambda _{\mathrm{x}%
}(t)\sigma _{\mathrm{s}}^{\mathrm{x}}+\lambda _{\mathrm{z}}(t)\sigma _{%
\mathrm{s}}^{\mathrm{z}}$, $\{\sigma ^{\mathrm{i}}\}$ are the Pauli
operators, $\lambda _{\mathrm{x}}(t)$ and $\lambda _{\mathrm{z}}(t)$ are two
time-dependent paramaters, and the spin bath Hamiltonian $H_{\mathrm{b}%
}=-\omega _{\mathrm{b}}\sum_{\mathrm{k}}\sigma _{\mathrm{k}}^{\mathrm{z}%
}-h\sum_{\left\langle \mathrm{k,j}\right\rangle }\sigma _{\mathrm{k}}^{%
\mathrm{x}}\sigma _{\mathrm{j}}^{\mathrm{x}}$ is that of an Ising chain.

So what kind of original composite system Hamiltonian $\mathcal{H}(t)$ can
be mapped to the $H_{\mathrm{uc}}$ specified in Eq.~(\ref{s9}) above? To
that end we first show how to derive the following relation:%
\begin{equation}
\exp (ig\Sigma ^{\mathrm{x}}J^{\mathrm{x}})J^{\mathrm{z}}\exp (-ig\Sigma ^{%
\mathrm{x}}J^{\mathrm{x}})=J^{\mathrm{z}}\cos (g\Sigma ^{\mathrm{x}})+J^{%
\mathrm{y}}\sin (g\Sigma ^{\mathrm{x}}),  \label{S6}
\end{equation}%
where $\{\Sigma ^{\mathrm{x}}\}$ and $\{J^{\mathrm{z}}\}$, respectively, are
the angular momentum operators for two different spin systems along
directions $x$, $z$ etc. To otain this relation, we first use the
Baker-Hausdorff lemma such that the left-hand side of the above equation can
be written as
\begin{eqnarray}
\exp (ig\Sigma ^{\mathrm{x}}J^{\mathrm{x}})J^{\mathrm{z}}\exp (-ig\Sigma ^{%
\mathrm{x}}J^{\mathrm{x}}) &=&J^{\mathrm{z}}+[ig\Sigma ^{\mathrm{x}}J^{%
\mathrm{x}},J^{\mathrm{z}}]+\frac{1}{2!}[ig\Sigma ^{\mathrm{x}}J^{\mathrm{x}%
},[ig\Sigma ^{\mathrm{x}}J^{\mathrm{x}},J^{\mathrm{z}}]]  \notag \\
&&+\frac{1}{3!}[ig\Sigma ^{\mathrm{x}}J^{\mathrm{x}},[ig\Sigma ^{\mathrm{x}%
}J^{\mathrm{x}},[ig\Sigma ^{\mathrm{x}}J^{\mathrm{x}},J^{\mathrm{z}%
}]]]+\cdots .
\end{eqnarray}%
With the commutators of angular momentum, it can be simplified to
\begin{gather}
\exp (ig\Sigma ^{\mathrm{x}}J^{\mathrm{x}})J^{\mathrm{z}}\exp (-ig\Sigma ^{%
\mathrm{x}}J^{\mathrm{x}})=J^{\mathrm{z}}+g\Sigma ^{\mathrm{x}}J^{\mathrm{y}%
}-\frac{1}{2!}(g\Sigma ^{\mathrm{x}})^{2}J^{\mathrm{z}}-\frac{1}{3!}(g\Sigma
^{\mathrm{x}})^{3}J^{\mathrm{y}}+\cdots  \notag \\
=J^{\mathrm{z}}\cos (g\Sigma ^{\mathrm{x}})+J^{\mathrm{y}}\sin (g\Sigma ^{%
\mathrm{x}}),
\end{gather}%
which is just the relation we hope to derive.

Applying the relation of Eq.~(\ref{S6}) to the expressions $\exp (igG)\sigma
_{\mathrm{s}}^{\mathrm{z}}\exp (-igG)$ and $\exp (igG)\sigma _{\mathrm{k}}^{%
\mathrm{z}}\exp (-igG)$ with $G=\sigma _{\mathrm{s}}^{\mathrm{x}}\sum_{%
\mathrm{k}}\sigma _{\mathrm{k}}^{\mathrm{x}}$, we obtain
\begin{eqnarray*}
\mathcal{H}(t) &=&\lambda _{\mathrm{x}}(t)\sigma _{\mathrm{s}}^{\mathrm{x}%
}-\cos (g)\omega _{\mathrm{b}}\sum_{\mathrm{k}}\sigma _{\mathrm{k}}^{\mathrm{%
z}}-h\sum_{\left\langle \mathrm{k,j}\right\rangle }\sigma _{\mathrm{k}}^{%
\mathrm{x}}\sigma _{\mathrm{j}}^{\mathrm{x}} \\
&&+\lambda _{\mathrm{z}}(t)[\sigma _{\mathrm{s}}^{\mathrm{z}}\cos (g\sum_{%
\mathrm{k}}\sigma _{\mathrm{k}}^{\mathrm{x}})+\sigma _{\mathrm{s}}^{\mathrm{y%
}}\sin (g\sum_{\mathrm{k}}\sigma _{\mathrm{k}}^{\mathrm{x}})]-\omega _{%
\mathrm{b}}\sin (g)\sigma _{\mathrm{s}}^{\mathrm{x}}\sum_{\mathrm{k}}\sigma
_{\mathrm{k}}^{\mathrm{y}},
\end{eqnarray*}%
which can be rewritten as $\mathcal{H}(t)=\mathcal{H}_{\mathrm{s}}+\mathcal{H%
}_{\mathrm{I}}+\mathcal{H}_{\mathrm{b}}$, with
\begin{eqnarray}
\mathcal{H}_{\mathrm{s}} &=&\lambda _{\mathrm{x}}(t)\sigma _{\mathrm{s}}^{%
\mathrm{x}}+\lambda _{\mathrm{z}}(t)\sigma _{\mathrm{s}}^{\mathrm{z}},
\notag \\
\mathcal{H}_{\mathrm{b}} &=&-\cos (g)\omega _{\mathrm{b}}\sum_{\mathrm{k}%
}\sigma _{\mathrm{k}}^{\mathrm{z}}-h\sum_{\left\langle \mathrm{k,j}%
\right\rangle }\sigma _{\mathrm{k}}^{\mathrm{x}}\sigma _{\mathrm{j}}^{%
\mathrm{x}},  \notag \\
\mathcal{H}_{\mathrm{I}} &=&-\omega _{\mathrm{b}}\sin (g)\sigma _{\mathrm{s}%
}^{\mathrm{x}}\sum_{\mathrm{k}}\sigma _{\mathrm{k}}^{\mathrm{y}}+\lambda _{%
\mathrm{z}}(t)\sigma _{\mathrm{s}}^{\mathrm{z}}[\cos (g\sum_{\mathrm{k}%
}\sigma _{\mathrm{k}}^{\mathrm{x}})-1]+\lambda _{\mathrm{z}}(t)\sigma _{%
\mathrm{s}}^{\mathrm{y}}\sin (g\sum_{\mathrm{k}}\sigma _{\mathrm{k}}^{%
\mathrm{x}}).
\end{eqnarray}%
If the coupling strength $g$ is sufficiently small, one may Taylor expand
the sine and cosine functions in the expression for $\mathcal{H}_{\mathrm{I}}
$ above, reducing it to $\mathcal{H}_{\mathrm{I}}=-g\omega _{\mathrm{b}%
}\sigma _{\mathrm{s}}^{\mathrm{x}}\sum_{\mathrm{k}}\sigma _{\mathrm{k}}^{%
\mathrm{y}}+g\lambda _{\mathrm{z}}(t)\sigma _{\mathrm{s}}^{\mathrm{y}}\sum_{%
\mathrm{k}}\sigma _{\mathrm{k}}^{\mathrm{x}}$. Summarizing our explicit
calculations above as an example, the key message is that an open quantum
system with $\mathcal{H}_{\mathrm{s}}$, $\mathcal{H}_{\mathrm{b}}$, and $%
\mathcal{H}_{\mathrm{I}}$ as above can indeed be mapped onto a system-bath
decoupled system $H_{\mathrm{uc}}$ through the generator $G=\sigma _{\mathrm{%
s}}^{\mathrm{x}}\sum_{\mathrm{k}}\sigma _{\mathrm{k}}^{\mathrm{x}}$. Indeed,
with this \textquotedblleft reversed engineering" approach, it is clear that
a wide variety of mapping between an original quantum open system and $H_{%
\mathrm{uc}}$ can be generated by choosing different $G$ as the mapping
generators. The existence of this kind of mapping within such a class of
problems indicates that we can push thermodynamics concepts to the strong
SBC regime without any conceptual difficulty.

As a second example we now discuss a Caldeira-Leggett like model that can be
mapped onto a system-bath decoupled model $H_{\mathrm{uc}}=H_{b}+H_{\mathrm{s%
}}(t)$, with $H_{b}=\sum_{k}\omega _{k}a_{k}^{\dagger }a_{k}$, and $H_{%
\mathrm{s}}(t)=\frac{p_{\mathrm{s}}^{2}}{2m}+\frac{1}{2}\lambda _{t}{x_{%
\mathrm{s}}}^{2}$, where $\lambda _{t}$ a time-depenndent parameter due to
the change of the trapping frequency in a work protocol. For future
reference we also define $q_{k}=(a_{k}^{\dagger }+a_{k})$ and $\pi
_{k}=i(a_{k}^{\dagger }-a_{k})$. Consider then again the transformation $%
\mathcal{H}(t)=e^{igG}H_{\mathrm{uc}}(t)e^{-igG}$, now with $G=p_{\mathrm{s}%
}\sum g_{k}(a_{k}^{\dagger }+a_{k})$. Through the property of displacement
operator $D(-\alpha )aD(\alpha )=a+\alpha $\ with $D(\alpha )=\exp (\alpha
a^{\dagger }-\alpha ^{\ast }a)$, we obtain
\begin{equation}
e^{igG}(\sum \omega _{k}a_{k}^{\dagger }a_{k})e^{-igG}=\sum \omega
_{k}(a_{k}^{\dagger }-igg_{k}p_{\mathrm{s}})(a_{k}+igg_{k}p_{\mathrm{s}}),
\end{equation}%
and
\begin{equation}
e^{igG}[\frac{1}{2m}p_{\mathrm{s}}^{2}+V(\lambda _{t},x_{\mathrm{s}%
})]e^{-igG}=\frac{1}{2m}p_{\mathrm{s}}^{2}+V(\lambda _{t},x_{\mathrm{s}}-%
\sqrt{2}g\sum g_{k}q_{k}),
\end{equation}%
where $V(\lambda _{t},x)=\frac{1}{2}\lambda _{t}x^{2}$ is the potential
function. Combining these two equations, the original system Hamiltonian $%
\mathcal{H}_{s}$ of a Caldeira-Leggett like model is found to be
\begin{equation}
\mathcal{H}_{\mathrm{s}}=\frac{1}{2M_{\mathrm{s}}}p_{\mathrm{s}%
}^{2}+V(\lambda _{t},x_{\mathrm{s}}),
\end{equation}%
where $M_{\mathrm{s}}=m/(1+$ $2mg^{2}\sum_{\mathrm{k}}\omega _{\mathrm{k}}g_{%
\mathrm{k}}^{2})$ is the renormalized mass. The orginal $\mathcal{H}_{%
\mathrm{b}}$ before the mapping also differs from the $H_{\mathrm{b}}$ after
the mapping, namely,
\begin{equation}
\mathcal{H}_{\mathrm{b}}=\sum_{\mathrm{k}}\omega _{\mathrm{k}}a_{\mathrm{k}%
}^{\dagger }a_{\mathrm{k}}+\frac{1}{2}\lambda _{\mathrm{t}}(g\sum_{\mathrm{k}%
}g_{\mathrm{k}}q_{\mathrm{k}})^{2}.
\end{equation}%
Finally, the original system-bath coupling term before the mapping:
\begin{equation}
\mathcal{H}_{\mathrm{I}}=g\lambda _{\mathrm{t}}x_{\mathrm{s}}\sum_{\mathrm{k}%
}g_{\mathrm{k}}q_{\mathrm{k}}+gp_{\mathrm{s}}\sum_{\mathrm{k}}g_{\mathrm{k}%
}\pi _{\mathrm{k}},
\end{equation}%
where the first term is the standard interaction Hamiltonian adopted in the
Caldeira-Leggett model, with $gg_{\mathrm{k}}$ depicting the coupling
strength between the system and the $\mathrm{k}$th harmonic oscillator. This
second working example further illustrates that it is feasible to connect an
original open quantum system Hamiltonian $\mathcal{H}(t)$ with a system-bath
decoupled Hamiltonian $H_{\mathrm{uc}}(t)$ through a mapping. The existence
of a mapping in such class of problems makes it possible to work on quantum
thermodynamic quantities rigorously and identify the underlying fluctuation
theorems, as shown in the main text and later sections here.

To complement the simulation results presented in the main text using the
spin model, in this supplementary material, we shall also use the above
Caldeira-Leggett like model to further illustrate our results. Specifically,
we consider a simple physical situation motivated by the specific physical
context of sympathetic cooling of atomic gases \cite{sympathetic cooling}.
Consider then Cs-Li atoms \cite{Mudrich} [likewise, K-Rb atoms studied in
Ref.~\cite{Modugno}] trapped in an optical dipole trap \cite{Mudrich}
approximated by harmonic potentials \cite{Engler}. In our example, we take a
lithium atom as system and a cesium atom as the bath. To further simplify
the problem, we only consider one of the radial oscillations. The composite
system is then reduced to one harmonic oscillator interacting with a single
free oscillator as the bath, with $H_{\mathrm{b}}=\omega _{\mathrm{Cs}}^{%
\mathrm{r}}a^{\dagger }a$. The system Hamiltonian is taken as $H_{\mathrm{s}%
}(t)=\frac{p_{\mathrm{s}}^{2}}{2m}+\frac{1}{2}m\omega _{\mathrm{s}%
}^{2}(t)x^{2}$ with $\omega _{\mathrm{s}}(t)=\omega _{\mathrm{Li}}^{\mathrm{r%
}}(1+\alpha t)$ and the trapping-frequency ramping rate $\alpha $. According
to our discussions above, the expression of $\mathcal{H}(t)$ mapped to such $%
H_{\mathrm{uc}}=H_{\mathrm{s}}(t)+H_{\mathrm{b}}$ can be easily found if we
use $G=p_{\mathrm{s}}g(a^{\dagger }+a)$.

\section{Quantum thermodynamics with weak system-bath coupling}

To appreciate the treatment we put forward in the main text to handle open
quantum systems with strong SBC, it is necessary to recap some known results
of quantum thermodynamics under the assumption of weak SBC. To consider
cases with vanishingly small SBC, it is convinient to simply use $H_{\mathrm{%
s}}$ and $H_{\mathrm{b}}$ to denote the system and bath Hamiltonians
respectively, the notation deliberately chosen to be the same as what is
used in the main text when considering the uncoupled Hamiltonian in the
three-stage time evolution picture. We also use $H_{\mathrm{uc}}=H_{\mathrm{s%
}}+H_{\mathrm{b}}$ to represent the whole system-bath Hamiltonian, upon
neglecting their coupling. The system and bath as a whole isolated system
will be thermalized to the following equilibrium state $\Omega _{\mathrm{uc}%
} =\frac{1}{Z_\mathrm{uc}}\exp (-\beta H_{\mathrm{uc}})$, where $Z_{\mathrm{u%
}c}=\mathrm{Tr}[\exp (-\beta {H}_\mathrm{uc})]$ is the partition function, $%
\beta =1/T$ is the inverse temperature of a super bath, and the Boltzmann
constant is set to $k_{B}=1$. Since we have assumed vanishingly small weak
SBC, this equilibrium state results in the system's equilibrium reduced
state, $\varpi _{\mathrm{s}} =\exp (-\beta H_{{s}})/Z_{\mathrm{s}}$, where $%
Z_{\mathrm{s}}$ is the associated partition function of $H_{\mathrm{s}}$.

Under weak SBC, the basic quantum thermodynamic quantities can be defined as
follows. First, one may define work as \cite{Alicki}
\begin{equation}
dW_{\mathrm{w}}=\mathrm{Tr}_{\mathrm{s}}(\rho _{\mathrm{s}}\ dH_{\mathrm{s}%
}),  \label{Wweak}
\end{equation}%
where $\rho _{\mathrm{s}}=\mathrm{Tr}_{\mathrm{b}}(\rho )$ and $\rho $ are
the density matrices of the system and the overall system-bath as a whole
system, respectively. Here and later we use the subscript or supercript
\textquotedblleft w\textquotedblright\ to highlight that we are working with
the conventional work and heat quantities under weak SBC. This definition of
work has been applied to both the equilibrium thermodynamics and
non-equilibrium processes because it is fully consistent with the well-known
two-time energy measurement operation definition if the initial state does
not have coherence between energy eigenstates.

On the other hand, the change of the total internal energy of the system can
be captured by
\begin{equation}
dE=d\mathrm{Tr}_{\mathrm{s}}(\rho _{\mathrm{s}}H_{\mathrm{s}}).
\end{equation}%
Consider now a process where $dW_{\mathrm{w}}=0$, then all the internal
energy change should be due to heat exchange. That is, $dQ_{\mathrm{w}}=dE$
with $\mathrm{Tr}_{\mathrm{s}}(\rho _{\mathrm{s}}\ dH_{\mathrm{s}})=0$. This
indicates that
\begin{equation}
dQ_{\mathrm{w}}=\mathrm{Tr}_{\mathrm{s}}(H_{\mathrm{s}}\ d\rho _{\mathrm{s}%
}).
\end{equation}

One can further proceed to define the free energy and the thermodynamic
entropy, respectively:
\begin{eqnarray}
F_{\mathrm{w}} &=&-T\ln Z_{\mathrm{s}},  \notag \\
S_{\mathrm{w}} &=&S_{\mathrm{v}}(\varpi _{\mathrm{s}}),
\end{eqnarray}%
where $S_{\mathrm{v}}(\rho _{\mathrm{s}})=-\mathrm{Tr}_{\mathrm{s}}(\rho _{%
\mathrm{s}}\ln \rho _{\mathrm{s}})$ is the von Neumann entropy of the
equilibrium state $\rho _{\mathrm{s}}$. That is, for equilbirum situations
with weak SBC, the thermodynamics entropy equals to the von Neumann entropy
of the equilibrium state $\varpi _{\mathrm{s}}$. One can confirm this result
by further using the relation $dS_{\mathrm{w}}=\beta dQ_{\mathrm{w}}$. Using
these treatments, other thermodynamics relations such as $F_{\mathrm{w}%
}=E-TS_{\mathrm{w}}$ and $S_{\mathrm{w}}=-[\partial F_{\mathrm{w}}/\partial
T]_{V}$ can be all recovered.

For later discussions, it is also necessary to first discuss known quantum
fluctuation theorems with weak SBC. To that end let us focus on the seminal
quantum Tasaki-Crooks fluctuation theorem \cite{two}. We assume that the
system is isolated from the bath when work is done during the time interval
of $[t_0, t]$ and that work done can be measured through the well-known
two-time energy measurement protocol. The fluctuation of work can be fully
characterized by the work probability density $p_{t_{0}\rightarrow t}^{%
\mathrm{w}}(w)$. The so-called characteristic function of work can be
defined as the Fourier transform of the work probability density $%
p_{t_{0}\rightarrow t}^{\mathrm{w}}(w)$, namely,
\begin{equation}
\Theta _{t_{0}\rightarrow t}^{\mathrm{w}}(u)=\int
dwe^{iuw}p_{t_{0}\rightarrow t}^{\mathrm{w}}(w).
\end{equation}%
As explicitly shown in \cite{dtcf}, this characteristic function is
connected with the system's time evolution operators through
\begin{equation}
\Theta _{t_{0}\rightarrow t}^{\mathrm{w}}(u)=Z_{\mathrm{s}}(t_0)^{-1}\mathrm{%
Tr}_{\mathrm{s}}[U_{t_{0}\rightarrow t}^{\mathrm{s}\dagger }e^{iuH_{\mathrm{s%
}}(t)}U_{t_{0}\rightarrow t}^{\mathrm{s}}e^{-iuH_{\mathrm{s}%
}(t_{0})}e^{-\beta H_{\mathrm{s}}(t_{0})}],
\end{equation}%
where the partition function of the system is $Z_{\mathrm{s}}(t_{0})=\mathrm{%
Tr}_{\mathrm{s}}[e^{-\beta H_{\mathrm{s}}(t_{0})}]$, and the time evolution
operator of system is $U_{t_{0}\rightarrow t}^{\mathrm{s}}=\mathcal{T}\exp
[-i\int_{t_{0}}^{t}H_{\mathrm{s}}(s)ds]$. If a complex parameter $%
v=-u+i\beta $ is used, the above work characteristic function can be further
transformed to%
\begin{equation}
\Theta _{t_{0}\rightarrow t}^{\mathrm{w}}(u)=Z_{\mathrm{s}}(t_{0})^{-1}%
\mathrm{Tr}_{\mathrm{s}}[e^{-ivH_{\mathrm{s}}(t)}e^{-\beta H_{\mathrm{s}%
}(t)}U_{t_{0}\rightarrow t}^{\mathrm{s}}e^{ivH_{\mathrm{s}%
}(t_{0})}U_{t_{0}\rightarrow t}^{\mathrm{s}\dagger }].
\end{equation}%
Since the system is isolated from its bath, its time evolution is reversible
(i.e., $U_{t_{0}\rightarrow t}^{\mathrm{s}}=U_{t\rightarrow t_{0}}^{\mathrm{s%
}\dagger }$). Using this reversibility, we obtain
\begin{eqnarray}
\Theta _{t_{0}\rightarrow t}^{\mathrm{w}}(u) &=&Z_{\mathrm{s}}(t_{0})^{-1}Z_{%
\mathrm{s}}(t)Z_{\mathrm{s}}(t)^{-1}\mathrm{Tr}_{\mathrm{s}}[U_{t\rightarrow
t_{0}}^{\mathrm{s}\dagger }e^{ivH_{\mathrm{s}}(t_{0})}U_{t\rightarrow
t_{0}}^{\mathrm{s}}e^{-ivH_{\mathrm{s}}(t)}e^{-\beta H_{\mathrm{s}}(t)}]
\notag \\
&=&Z_{\mathrm{s}}(t_{0})^{-1}Z_{\mathrm{s}}(t)\Theta _{t\rightarrow t_{0}}^{%
\mathrm{w}}(v)  \notag \\
&=&Z_{\mathrm{s}}(t_{0})^{-1}Z_{\mathrm{s}}(t)\Theta _{t\rightarrow t_{0}}^{%
\mathrm{w}}(-u+i\beta ),
\end{eqnarray}%
where $\Theta _{t\rightarrow t_{0}}^{\mathrm{w}}(v)$ represents the
characteristic function of work in the time-reversed process with the
Hamiltonian of the system driven from $H_{\mathrm{s}}(t)\rightarrow H_{%
\mathrm{s}}(t_{0})$. The quantum Tasaki-Crooks fluctuation theorem is then
obtained below by use of the inverse Fourier transform, namely,
\begin{equation}
\frac{p_{t_{0}\rightarrow t}^{\mathrm{w}}(w)}{p_{t\rightarrow t_{0}}^{%
\mathrm{w}}(-w)}=\frac{Z_{\mathrm{s}}(t)}{Z_{\mathrm{s}}(t_{0})}e^{\beta w}.
\label{JZ2}
\end{equation}

By multiplying $p_{t\rightarrow t_{0}}^{\mathrm{w}}(-w)$ to both sides of
Eq.~(\ref{JZ2}) and then integrating both sides, one directly obtain the
celebrated Jarzynski equality
\begin{equation}
\left\langle e^{-\beta w_{\mathrm{w}}}\right\rangle =e^{-\beta \Delta F_{%
\mathrm{w}}},
\end{equation}%
with $\left\langle e^{-\beta w_{\mathrm{w}}}\right\rangle =\int
p_{t_{0}\rightarrow t}^{\mathrm{w}}(w)e^{-\beta w}dw$.

\begin{figure}[tbph]
\centering\includegraphics[width=5.5in]{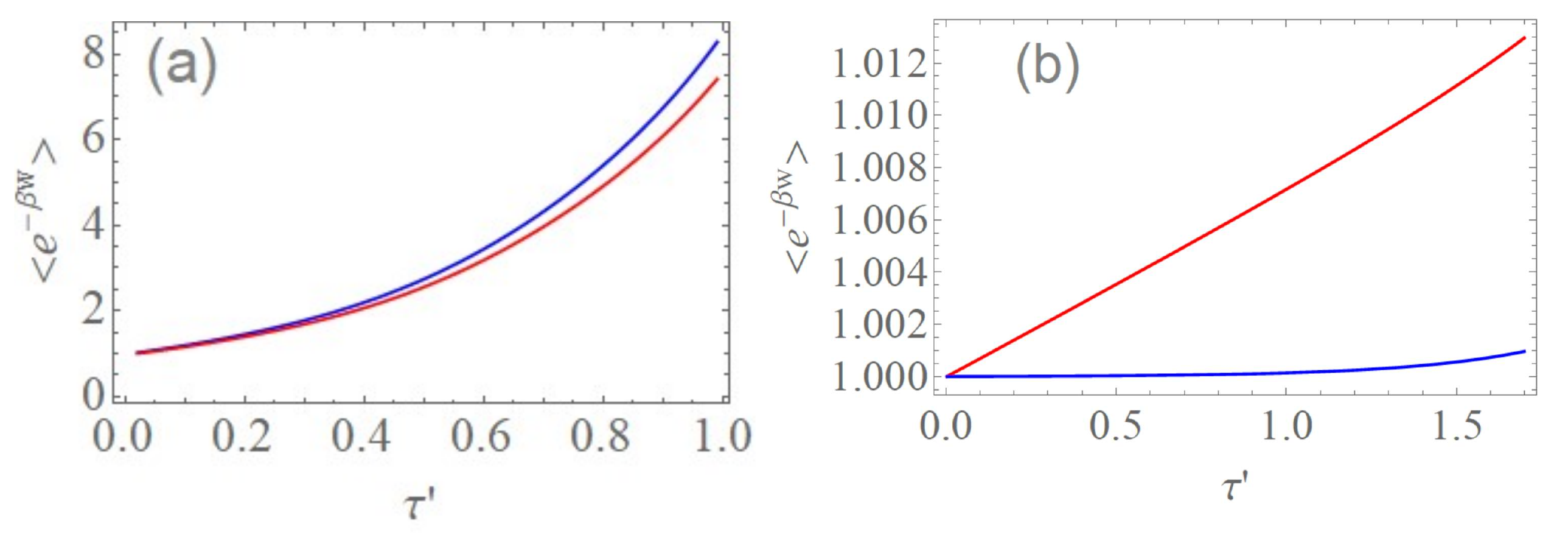}
\caption{Comparison between the quantities $\left\langle e^{-\protect\beta %
w_{\mathrm{w}}}\right\rangle $ with non-zero coupling (red line) and $e^{-%
\protect\beta \Delta F_{\mathrm{w}}}$ (blue line), as a function of the work
protocol duration $\protect\tau ^{\prime }$, with the initial state chosen
as the Gibbs state $\Omega $. (a) shows the results in the central spin
model, and the parameters are inverse temperature $\protect\beta =1$, $%
\protect\lambda _{\mathrm{x}}(0)=1$, $\protect\lambda _{\mathrm{z}}(0)=1$, $%
\protect\omega _{\mathrm{b}}=1$, $h=1$, $\protect\alpha _{\mathrm{x}}=1$, $%
\protect\alpha _{\mathrm{z}}=-0.6$, coupling strength $g=0.3$ and number of
bath spins $N=6$. (b) shows the results in the Caldeira-Leggett like model,
and the parameters are chosen as inverse temperature $\protect\beta =4.08$,
trapping-frequency of bath $\protect\omega _{\mathrm{Cs}}^{\mathrm{r}}=1$,
trapping-frequency of system $\protect\omega _{\mathrm{Li}}^{\mathrm{r}}=2.83
$, trapping-frequency ramping rate $\protect\alpha =-0.1$ and coupling
strength $g=0.1$.}
\label{Figs1}
\end{figure}

To conclude this section, it must be stressed that in cases of strong SBC,
one can no longer just compute the reduced state of the system and plug it
into the above expressions of $dW_{\mathrm{w}}$ and $dQ_{\mathrm{w}}$ to do
quantum thermodynamics. Indeed, as shown in Fig.~\ref{Figs1}, simply doing
so will lead to serious problems/inconsistencies, including a violation of
the celebrated work fluctuation theorems. Indeed, the work defined in Eq.~(%
\ref{Wweak}) is invalid at non-zero SBC strength, especially when the
duration of the protocol is not short. In Fig.~\ref{Figs1}, it is seen that
as the duration of the work protocol increases, the deviation of $%
\left\langle e^{-\beta w_{\mathrm{w}}}\right\rangle$ from $e^{-\beta \Delta
F_{\mathrm{w}}}$ becomes more obvious, indicating that the second law of
thermodynamics may not be respected if people nailvely applies the concept
of work for weak SBC to situations with strong SBC. These results also imply
the necessity of an alternative approach to the calculation of work done in
the presence of strong SBC.

\section{Work fluctuation theorems with strong system-bath coupling}

The three-stage time evolution picture makes it possible to know the
exientence of work as a definite thermodynamic quantity, even when there is
strong SBC. In particular, with the decomposition of the original time
evolution operator into the products of three terms as shown in Eq.~(S1) or
Eq.~(\ref{w0neq}), work done can be entirely associated with the propagator $%
U_{\mathrm{uc}}$, which depicts a system without its coupling to the bath.
Hence, the work with strong SBC can be investigated by checking the work
done associated with $U_{\mathrm{uc}}$ for the time interval interval $%
[0,\tau ^{\prime }]$. As also mentioned in the main text, we assume that the
initial state of the work process is the Gibbs state $\Omega _{\mathrm{uc}}$
(as the result of a long self-relaxation process) and so the initial state
of the system is$\ \varpi _{\mathrm{s}}(0)$, defined in the previous
section, with no initial coherence between the energy eigenstates. The
characteristic function of work will be exactly the same as that computed
from $U_{\mathrm{uc}}$, namely, $\Theta _{0\rightarrow \tau ^{\prime }}^{%
\mathrm{w}}(u)$. Using the expessions of work characteristic functions in
the previous section and also adding the time-independent bath Hamiltinian $%
H_{\mathrm{b}}$, we obtain
\begin{equation}
\Theta _{0\rightarrow \tau ^{\prime }}^{\mathrm{w}}(u)=Z(0)^{-1}\mathrm{Tr}%
\{U_{\mathrm{uc}}^{\dagger }e^{iu[H_{\mathrm{s}}(\tau ^{\prime })+H_{\mathrm{%
b}}]}U_{\mathrm{uc}}e^{-iu[H_{\mathrm{s}}(0)+H_{\mathrm{b}}]}e^{-\beta
\lbrack H_{\mathrm{s}}(0)+H_{\mathrm{b}}]}\},
\end{equation}%
where the commutator $[U_{\mathrm{uc}},H_{\mathrm{b}}]=0$, $U_{\mathrm{uc}%
}=U_{t_{0}\rightarrow \tau ^{\prime }}^{\mathrm{s}}e^{-iH_{\mathrm{b}}\tau
^{\prime }}$\ and $Z(0)=Z_{\mathrm{s}}(0)Z_{\mathrm{b}}$\ with $Z_{\mathrm{b}%
}=\mathrm{Tr}_{\mathrm{b}}e^{-\beta H_{\mathrm{b}}}$\ are used. Further
using the relation $U_{\mathrm{uc}}=e^{-igG}U_{0\rightarrow \tau ^{\prime
}}e^{igG}$, the the above equation can be transformed to
\begin{eqnarray}
\Theta _{t_{0}\rightarrow \tau ^{\prime }}^{\mathrm{w}}(u) &=&Z(0)^{-1}%
\mathrm{Tr}\{e^{-igG}U_{0\rightarrow \tau ^{\prime }}^{\dagger
}e^{igG}e^{iu[H_{\mathrm{s}}(\tau ^{\prime })+H_{\mathrm{b}%
}]}e^{-igG}U_{0\rightarrow \tau ^{\prime }}e^{igG}e^{-iu[H_{\mathrm{s}%
}(0)+H_{\mathrm{b}}]}e^{-\beta \lbrack H_{\mathrm{s}}(0)+H_{\mathrm{b}}]}\}
\notag \\
&=&Z(0)^{-1}\mathrm{Tr}\{U_{0\rightarrow \tau ^{\prime }}^{\dagger
}e^{igG}e^{iu[H_{\mathrm{s}}(\tau ^{\prime })+H_{\mathrm{b}%
}]}e^{-igG}U_{0\rightarrow \tau ^{\prime }}e^{igG}e^{-iu[H_{\mathrm{s}%
}(0)+H_{\mathrm{b}}]}e^{-igG}e^{igG}e^{-\beta \lbrack H_{\mathrm{s}}(0)+H_{%
\mathrm{b}}]}e^{-igG}\}.  \notag \\
&&
\end{eqnarray}%
Through the relation $\mathcal{H}(t)=e^{igG}[H_{\mathrm{s}}(t)+H_{\mathrm{b}%
}]e^{-igG}$, the above characteristic function is then reduced to
\begin{equation}
\Theta _{0\rightarrow \tau ^{\prime }}(u)=Z(0)^{-1}\mathrm{Tr}%
[U_{0\rightarrow \tau ^{\prime }}^{\dagger }e^{iu\mathcal{H}(\tau ^{\prime
})}U_{0\rightarrow \tau ^{\prime }}e^{-iu\mathcal{H}(0)}e^{-\beta \mathcal{H}%
(0)}].
\end{equation}%
Note that this expression above for work characteristic function is now
entirely about the original system-bath Hamiltonian $\mathcal{H}$, and hence
we have arrived at a key expression for work fluctuation theorems with
strong SBC. For this reason, in the equation above we have also dropped the
\textquotedblleft w\textquotedblright\ superscript.

\section{Work, heat and internal energy at arbitrary SBC strength}

As outlined in the main text and also in previous sections, the heat
exchanged in the original system with arbitrary SBC strength during the time
interval $[0,\tau^{\prime}]$ can be calculated from $U_{\mathrm{%
w\tau^{\prime}}}$ and $U_{\mathrm{w0}}$, the two auxiliary evolution
operators associated with time intervals $[-\tau ,0]$ and $[\tau ^{\prime
},\tau +\tau ^{\prime }]$, where the system-bath coupling can be understood
to be vanishingly small. To focus on the differential behaviors of heat
exchange, let us also assume $\tau ^{\prime }$ to be infinitely small.

The total heat exchange hence occurs during the two self-relaxation time
intervals $[-\tau ,0]$ and $[\tau ^{\prime },\tau +\tau ^{\prime }]$. There
is no ambuity in calculating heat during these two processes because in
above we have already introduced how heat should be calculated under weak
SBC. The total heat exchange is then found to be the following:
\begin{eqnarray}
dQ &=&\int_{-\tau }^{0}\mathrm{Tr}_{\mathrm{s}}(H_{\mathrm{s}}\ d\rho _{%
\mathrm{s}})+\int_{\tau ^{\prime }}^{\tau ^{\prime }+\tau }\mathrm{Tr}_{%
\mathrm{s}}(H_{\mathrm{s}}\ d\rho _{\mathrm{s}})  \notag \\
&=&\mathrm{Tr}_{\mathrm{s}}[H_{\mathrm{s}}(-\tau )\rho _{\mathrm{s}}(0)]-%
\mathrm{Tr}_{\mathrm{s}}[H_{\mathrm{s}}(-\tau )\rho _{\mathrm{s}}(-\tau )]+%
\mathrm{Tr}_{\mathrm{s}}[H_{\mathrm{s}}(\tau ^{\prime })\rho _{\mathrm{s}%
}(\tau ^{\prime }+\tau )]-\mathrm{Tr}_{\mathrm{s}}[H_{\mathrm{s}}(\tau
^{\prime })\rho _{\mathrm{s}}(\tau ^{\prime })],
\end{eqnarray}%
where we have used, consistent with our construction, that the system $H_{%
\mathrm{s}}$ does not change during the time intervals $[-\tau ,0]$ and $%
[\tau ^{\prime },\tau +\tau ^{\prime }]$, always fixed at either $H_s(0)$ or
$H_s(\tau^{\prime})$. Further using the obvious relation $\mathrm{Tr}_{%
\mathrm{s}}[A_{\mathrm{s}}(t)\rho _{\mathrm{s}}(t)]=\mathrm{Tr}[A_{\mathrm{s}%
}(t)\rho (t)]$ with $A_{\mathrm{s}}$ being any system operator, we can
reexpress this expression for heat in terms of the following:
\begin{equation}
dQ=\mathrm{Tr}[H_{\mathrm{s}}(-\tau )\rho (0)]-\mathrm{Tr}[H_{\mathrm{s}%
}(-\tau )\rho (-\tau )]+\mathrm{Tr}[H_{\mathrm{s}}(\tau ^{\prime })\rho
(\tau ^{\prime }+\tau )]-\mathrm{Tr}[H_{\mathrm{s}}(\tau ^{\prime })\rho
(\tau ^{\prime })].
\end{equation}

Next we wish to connect the above expression for heat with system-bath's
orginal Hamiltonian $\mathcal{H}$. This can be partially done. We first use
the following relations:
\begin{eqnarray}
\rho (0)&=&e^{-igG}\rho(-\tau )e^{igG}, \\
\rho (\tau ^{\prime })&=&U_{\mathrm{uc}}e^{-igG}\rho (-\tau)e^{igG}U_{%
\mathrm{uc}}^{\dagger }, \\
\rho (\tau ^{\prime }+\tau)&=&e^{igG}U_{\mathrm{uc}}e^{-igG}\rho (-\tau
)e^{igG}U_{\mathrm{uc}}^{\dagger}e^{-igG}.
\end{eqnarray}
In addition, for an infinitesimal $\tau ^{\prime }$, the time evolution
operator$\ U_{\mathrm{uc}}\ $ can be rewritten to the first order of $\tau
^{\prime }$ as $U_{\mathrm{uc}}=\mathcal{T}\exp [-i\int_{0}^{\tau ^{\prime
}}H_{\mathrm{uc}}(s)ds]\approx 1-iH_{\mathrm{uc}}(0)\tau ^{\prime }=1-iH_{%
\mathrm{uc}}(-\tau )\tau ^{\prime }$. Putting all these considerations
together, we obtain
\begin{eqnarray}
dQ &=&\mathrm{Tr}[H_{\mathrm{s}}(-\tau )e^{-igG}\rho (-\tau )e^{igG}]-%
\mathrm{Tr}[H_{\mathrm{s}}(-\tau )\rho (-\tau )]  \notag \\
&&+\mathrm{Tr}\{H_{\mathrm{s}}(\tau ^{\prime })e^{igG}[1-iH_{\mathrm{uc}%
}(-\tau )\tau ^{\prime }]e^{-igG}\rho (-\tau )e^{igG}[1+iH_{\mathrm{uc}%
}(-\tau )\tau ^{\prime }]e^{-igG}\}  \notag \\
&&-\mathrm{Tr}\{H_{\mathrm{s}}(\tau ^{\prime })[1-iH_{\mathrm{uc}}(-\tau
)\tau ^{\prime }]e^{-igG}\rho (-\tau )e^{igG}[1+iH_{\mathrm{uc}}(-\tau )\tau
^{\prime }]\}.
\end{eqnarray}%
To the first order of the time $\tau ^{\prime }$, this expression of heat
can be rearranged to become
\begin{eqnarray}
dQ &=&\mathrm{Tr}\{H_{\mathrm{s}}(-i)[e^{igG}H_{\mathrm{uc}}(-\tau
)e^{-igG},\rho ]\tau ^{\prime }\}  \notag \\
&&+\mathrm{Tr}(\rho dH_{\mathrm{s}})-\mathrm{Tr}[e^{igG}(dH_{\mathrm{s}%
})e^{-igG}\rho ],
\end{eqnarray}%
where $H_{\mathrm{s}}=H_{\mathrm{s}}(-\tau )=H_{\mathrm{s}}(0)$, $dH_{%
\mathrm{s}}=H_{\mathrm{s}}(\tau ^{\prime })-H_{\mathrm{s}}(-\tau )=H_{%
\mathrm{s}}(\tau ^{\prime })-H_{\mathrm{s}}(0)$ and we have denoted the very
initial density of the whole system-bath system $\rho (-\tau )$ as $\rho$.
Noting that the infinitesimal state evolution under the whole system-bath
Hamiltonian $\mathcal{H=}e^{iG}H_{\mathrm{uc}}(-\tau )e^{-iG}$ is actually $%
d\rho =-i[\mathcal{H},\rho (-\tau )]\tau ^{\prime }$, we finally have
\begin{eqnarray}
dQ &=&\mathrm{Tr}(H_{\mathrm{s}}\ d\rho )+\mathrm{Tr}[\rho\ d(H_{\mathrm{s}%
}-e^{igG}H_{\mathrm{s}}e^{-igG})]  \notag \\
&=&\mathrm{Tr}(H_{\mathrm{s}}\ d\rho )+\mathrm{Tr}\{\rho\ d[H_{\mathrm{s}%
}-e^{igG}(H_{\mathrm{s}}+H_{\mathrm{b}})e^{-igG}]\}  \notag \\
&=&\mathrm{Tr}_{\mathrm{s}}(H_{\mathrm{s}}\ d\rho _{\mathrm{s}})+\mathrm{Tr}%
[\rho\ d(H_{\mathrm{s}}-\mathcal{H})],  \label{qEx}
\end{eqnarray}%
where the time-independent Hamiltonian of the bath $d(e^{igG}H_{\mathrm{b}%
}e^{-igG})=0$ is used, and $d\rho _{\mathrm{s}}=\mathrm{Tr}_{\mathrm{b}%
}(d\rho )$. This is the expression presented in the main text. Note that
this expression of heat is markedly different from the previous expression $%
dQ_{\mathrm{w}}$ under weak SBC. To obtain $dQ$, one not only needs to know
the change in the reduced density of the system, as what we normally need in
cases of weak SBC, but also need to examine the second term in Eq.~(\ref{qEx}%
) that involves $H_{\mathrm{s}}$.

Having derived the heat expression, we now turn to work, which can be
investigated by focusing on the time interval $[0,\tau ^{\prime }]$ in our
three-stage picture. During this stage, the system and bath is uncoupled.
Thus, the work done, denoted $dW$, can be calculated by examining the
internal energy change to the system $H_s$. That is,
\begin{eqnarray}
dW &=&\mathrm{Tr}_{\mathrm{s}}[\rho _{\mathrm{s}}(\tau ^{\prime })H_{\mathrm{%
s}}(\tau ^{\prime })]-\mathrm{Tr}_{\mathrm{s}}[\rho _{\mathrm{s}}(0)H_{%
\mathrm{s}}(0)]  \notag \\
&=&\mathrm{Tr}[\rho (\tau ^{\prime })H_{\mathrm{uc}}(\tau ^{\prime })]-%
\mathrm{Tr}[\rho (0)H_{\mathrm{uc}}(0)],
\end{eqnarray}%
where, consistent with our treatment, we have assumed that $H_{\mathrm{b}}$
as part of $H_\mathrm{uc}$ is time independent. As in treating heat exchange
above, let us now rewrite both the states $\rho (0)$ and $\rho (\tau
^{\prime })$ in terms of the initial state $\rho (-\tau )$ through $\rho
(0)=e^{-igG}\rho (-\tau )e^{igG}$ and $\rho (\tau ^{\prime })=U_{\mathrm{uc}%
}e^{-igG}\rho (-\tau )e^{igG}U_{\mathrm{uc}}^{\dagger }$. Furthe considering
an infinitemal $\tau^{\prime }$ to capture the differential work $dW$, we
safely use the expansion $U_{\mathrm{uc}}\approx 1-iH_{\mathrm{uc}}(0)\tau
^{\prime }$ to the first order of time $\tau ^{\prime }$. This leads us to
\begin{eqnarray}
dW &=&\mathrm{Tr}\{\rho (-\tau )e^{igG}[1+iH_{\mathrm{uc}}(0)\tau ^{\prime
}][H_{\mathrm{uc}}(0)+(\partial _{t}H_{\mathrm{uc}})\tau ^{\prime }][1-iH_{%
\mathrm{uc}}(0)\tau ^{\prime }]e^{-igG}\}  \notag \\
&&-\mathrm{Tr}[\rho (-\tau )e^{igG}H_{\mathrm{uc}}(0)e^{-igG}]  \notag \\
&=&\mathrm{Tr}[\rho (-\tau )e^{igG}(\partial _{t}H_{\mathrm{uc}})\tau
^{\prime }e^{-igG}],
\end{eqnarray}%
where $(\partial _{t}H_{\mathrm{uc}})\tau ^{\prime }=dH_{\mathrm{uc}}$
represents the infinitesimal change in the Hamiltonian $H_\mathrm{uc}$: $H_{%
\mathrm{uc}}(\tau ^{\prime })-H_{\mathrm{uc}}(0)$.\ Finally, using the
relation $\mathcal{H=}e^{igG}H_{\mathrm{uc}}e^{-igG}$, the $dW$ expression
reduces to
\begin{equation}
dW=\mathrm{Tr}(\rho\ d\mathcal{H}).  \label{dWeq}
\end{equation}
This is the expression presented in the main text.

It now becomes interesting to check what happens to the energy throughout
the whole process that is digested in terms of three stages. For convenience
we still assume $\tau ^{\prime }$ to be infinitesimal. The total internal
energy change, caused by heat exchange and work done, must be given by
\begin{equation}
dE=\mathrm{Tr}_{\mathrm{s}}[\rho _{\mathrm{s}}(\tau +\tau ^{\prime })H_{%
\mathrm{s}}(\tau +\tau ^{\prime })-\rho _{\mathrm{s}}(-\tau )H_{\mathrm{s}%
}(-\tau )].
\end{equation}%
With the obvious relation $\mathrm{Tr}_{\mathrm{s}}[H_{\mathrm{s}}(t)\rho _{%
\mathrm{s}}(t)]=\mathrm{Tr}[H_{\mathrm{s}}(t)\rho (t)]$, we can rewrite $dE$
as the following:
\begin{equation}
dE=\mathrm{Tr}[\rho (\tau +\tau ^{\prime })H_{\mathrm{s}}(\tau +\tau
^{\prime })-\rho (-\tau )H_{\mathrm{s}}(-\tau )].
\end{equation}%
Again, if we now rewrite $\rho (\tau +\tau ^{\prime })$ in terms of $\rho
(-\tau )$, with
\begin{equation}
\rho (\tau +\tau ^{\prime })=e^{igG}U_{\mathrm{uc}}e^{-igG}\rho (-\tau
)e^{igG}U_{\mathrm{uc}}^{\dagger }e^{-igG}\approx e^{igG}[1-iH_{\mathrm{uc}%
}(-\tau )\tau ^{\prime }]e^{-igG}\rho (-\tau )e^{igG}[1+iH_{\mathrm{uc}%
}(-\tau )\tau ^{\prime }]e^{-igG}
\end{equation}%
to the first order of $\tau ^{\prime }$, we arrive at
\begin{eqnarray}
dE &=&\mathrm{Tr}\{H_{\mathrm{s}}(\tau ^{\prime })e^{igG}[1-iH_{\mathrm{uc}%
}(-\tau )\tau ^{\prime }]e^{-igG}\rho (-\tau )e^{igG}[1+iH_{\mathrm{uc}%
}(-\tau )\tau ^{\prime }]e^{-igG}\}-\mathrm{Tr}[\rho (-\tau )H_{\mathrm{s}%
}(-\tau )]. \\
&=&\mathrm{Tr}\{[H_{\mathrm{s}}(\tau ^{\prime })-H_{\mathrm{s}}(0)]\rho
(-\tau )]\}-i\mathrm{Tr}\{H_{\mathrm{s}}(\tau ^{\prime })[e^{igG}H_{\mathrm{%
uc}}(-\tau )e^{-igG},\rho (-\tau )]\}\tau ^{\prime }  \notag \\
&=&\mathrm{Tr}(\rho \ dH_{\mathrm{s}})+\mathrm{Tr}(H_{\mathrm{s}}\ d\rho )
\notag
\end{eqnarray}%
Comparing with expressions of $dW$ and $dQ$ in Eq.~(\ref{dWeq}) and ~(\ref%
{qEx}), one immediately observes
\begin{equation*}
dE=dW+dQ.
\end{equation*}%
This can be understood as the manifestion of the first law of quantum
thermodynamics in our treatment.



In Fig. S2 (a), (b), (d) and (e) we compare the basic quantities between
weak and strong SBC for both the central spin model and Caldeira-Leggett
like model. Similar to a trend observed in Fig.~S1, it is seen that the
difference is more apparent when the duration of the work protocol
increases, namely, a larger $\tau ^{\prime }$.

\begin{figure}[tbph]
\centering\includegraphics[width=7in]{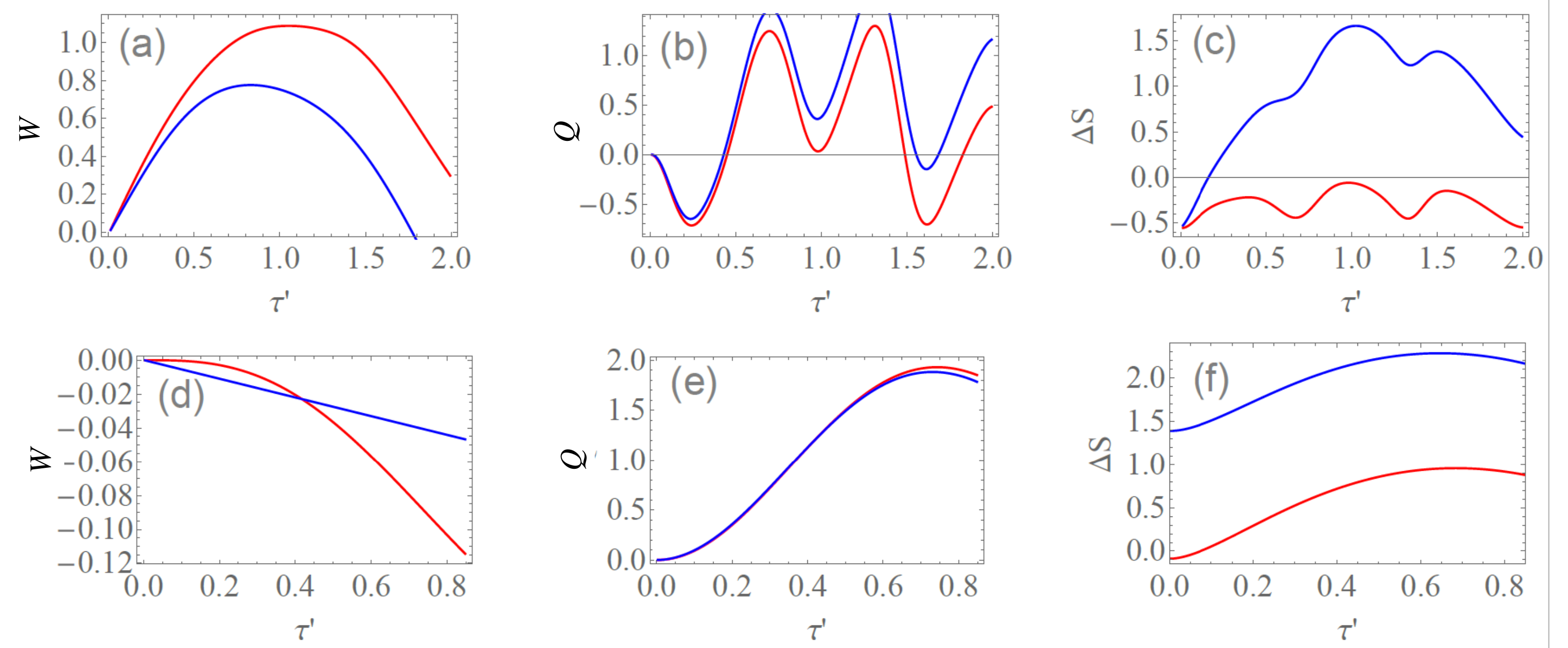}
\caption{Comparison in thermodynamics quantities between strong SBC (blue
lines) and weak SBC (red lines) as a function of the duration of the work
protocol $\protect\tau ^{\prime } $. (a), (b) and (c), respectively, show $W$%
, $Q$ and $\Delta S$ as a function of time for central spin model, and (d),
(e) and (f) are to compare the same but for the Caldeira-Leggett like model.
The corresponding parameters are the same as those in Fig.~S1.}
\end{figure}

\section{The entropy at strong coupling}

To obtain the entropy, first, we define thermodynamic entropy as $dS=\beta
dQ $, if and only if the process is quasi-static. Here as the presence of
bath, the quasi-static means that the composite system can always be
maintained in the global Gibbs state $\Omega $ \cite{Anders}. With the
definition of heat $dQ=\mathrm{Tr}(H_{\mathrm{s}}\ d\rho )+\mathrm{Tr}[\rho
\ d(H_{\mathrm{s}}-\mathcal{H})]$, we obtain
\begin{equation}
dS=-d\mathrm{Tr}(\Omega _{\mathrm{s}}\ln \varpi _{\mathrm{s}})-\mathrm{Tr}%
(\Omega d\ln \Omega )-d\ln \frac{Z_{\mathrm{s}}}{Z},
\end{equation}%
where the state of the composite system $\rho $ is replaced by the Gibbs
state $\Omega $, and the reduced state of the system is $\Omega _{\mathrm{s}%
}=\mathrm{Tr}_{\mathrm{b}}\Omega $. From the definition of partition
function the following relation $Z(t)=\mathrm{Tr}[e^{-\beta \mathcal{H}(t)}]=%
\mathrm{Tr}[e^{igG}e^{-\beta H_{\mathrm{uc}}(t)}e^{-igG}]=\mathrm{Tr}%
[e^{-\beta H_{\mathrm{uc}}(t)}]=Z_{\mathrm{s}}(t)Z_{\mathrm{b}}$ can be
obtained, and then the quantity $\frac{Z(t)}{Z_{\mathrm{s}}(t)}=Z_{\mathrm{b}%
}$ is time-independent. Then the last term is zero, i.e., $d\ln \frac{Z_{%
\mathrm{s}}}{Z}=0$. Also as $\mathrm{Tr}(\Omega d\ln \Omega )=d\mathrm{Tr}%
(\Omega )=0$, the entropy is reduced to
\begin{equation}
dS=-d\mathrm{Tr}(\Omega _{\mathrm{s}}\ln \varpi _{\mathrm{s}}).
\end{equation}%
With the definition of relative entropy, the change of thermodynamic entropy
is transformed to
\begin{equation}
dS=dS_{v}(\Omega _{\mathrm{s}})+dD(\left. \Omega _{\mathrm{s}}\right\vert
\varpi _{\mathrm{s}}),
\end{equation}%
where the von Neumann entropy is $S_{v}(\Omega _{\mathrm{s}})=-\mathrm{Tr}_{%
\mathrm{s}}(\Omega _{\mathrm{s}}\ln \Omega _{\mathrm{s}})$, and the relative
entropy is $D(\left. \Omega _{\mathrm{s}}\right\vert \varpi _{\mathrm{s}})=%
\mathrm{Tr}_{\mathrm{s}}[\Omega _{\mathrm{s}}\ln \Omega _{\mathrm{s}}]-%
\mathrm{Tr}_{\mathrm{s}}[\Omega _{\mathrm{s}}\ln \varpi _{\mathrm{s}}]$.
Fig. S2 (c) and (f) show the difference with that at weak coupling. Being
different from work and heat, the difference may arise at the very start of
the work protocol, because the entropy is a state variable.

\section{Details of the relation $\Delta F=\Delta E-T\Delta S$}

With the definition of free energy $F=T\ln Z_{\mathrm{s}}$ and the partition
function of system $Z_{\mathrm{s}}(t)=\mathrm{Tr}_{\mathrm{s}}\{\exp [-\beta
H_{\mathrm{s}}(t)]\}$, the change of free energy is%
\begin{equation}
dF=-T\ln Z_{\mathrm{s}}(\tau ^{\prime }+\tau )+T\ln Z_{\mathrm{s}}(-\tau ).
\end{equation}%
As mentioned in maintext, in both time intervals $[-\tau ,0$] and $[\tau
^{\prime },\tau +\tau ^{\prime }]$, heat is transferred while no work
protocol is executed via any change in the system parameters (given that the
Hamiltonians of the system is viewed as unchanged in these two time
intervals). Hence, the free energies are also unchanged in these time
intervals, and the free energy change can then be rewritten as
\begin{equation}
dF=-T\ln Z_{\mathrm{s}}(\tau ^{\prime })+T\ln Z_{\mathrm{s}}(0).
\end{equation}%
Since the Hamiltonian of the bath is time independent, its free energy is a
constant. As a result, the free energy change can be transformed to
\begin{eqnarray}
dF &=&-T\ln \mathrm{Tr}_{\mathrm{s}}\{e^{-\beta \lbrack H_{\mathrm{s}%
}(0)+dH_{\mathrm{s}}]}\}-T\ln \mathrm{Tr}_{\mathrm{b}}(e^{-\beta H_{\mathrm{b%
}}})+T\ln \mathrm{Tr}_{\mathrm{s}}[e^{-\beta H_{\mathrm{s}}(0)}]+T\ln
\mathrm{Tr}_{\mathrm{b}}(e^{-\beta H_{\mathrm{b}}})  \notag \\
&=&-T\ln \mathrm{Tr}\{e^{-\beta \lbrack H_{\mathrm{s}}(0)+dH_{\mathrm{s}}+H_{%
\mathrm{b}}]}\}+T\ln \mathrm{Tr}\{e^{-\beta \lbrack H_{\mathrm{s}}(0)+H_{%
\mathrm{b}}]}\},
\end{eqnarray}%
where time $\tau ^{\prime }$ is infinitely small, and the Hamiltonian is
rewritten as $H_{\mathrm{s}}(\tau ^{\prime })=H_{\mathrm{s}}(0)+dH_{\mathrm{s%
}}$. The free energy change is then reduced to
\begin{eqnarray}
dF &=&-T\ln \mathrm{Tr}\{e^{-\beta \lbrack H_{\mathrm{s}}(0)+H_{\mathrm{b}%
}]}(1-\beta dH_{\mathrm{s}})\}+T\ln \mathrm{Tr}\{e^{-\beta \lbrack H_{%
\mathrm{s}}(0)+H_{\mathrm{b}}]}\}  \notag \\
&=&\mathrm{Tr}[\varpi (0)d(H_{\mathrm{s}}+H_{\mathrm{b}})],
\end{eqnarray}%
where the time-independent Hamiltonian of the bath is used. Finally, we note
that $\varpi (0)=e^{-igG}\rho (-\tau )e^{igG}$ after assuming that the
initial state is the Gibbs state $\rho (-\tau )=\Omega (-\tau )$. As such,
we can further rewrite $dF$ as the following:
\begin{eqnarray}
dF &=&\mathrm{Tr}[e^{-igG}\rho (-\tau )e^{igG}d(H_{\mathrm{s}}+H_{\mathrm{b}%
})]  \notag \\
&=&\mathrm{Tr}(\rho \ d\mathcal{H}),
\end{eqnarray}%
where $\mathcal{H}=e^{igG}(H_{\mathrm{s}}+H_{\mathrm{b}})e^{-igG}$ is used.
This is consistent with our work expression since during a quasi-static
process at constant temperature, $dF$ is expected to be the same as $dW$.
Finally, combining with Eqs (S42) and (S47) the relation $\Delta F=\Delta
E-T\Delta S$ mentioned in the main text can be obtained.


\begin{thebibliography}{99}
\bibitem{quenching} The interaction Hamiltonian is quenched at the time $t=0$
and $t=\tau ^{\prime }$, and then there are some work added for finite time
interval $\tau $. These two work will approach to zero in the limit $\tau
\rightarrow +\infty $, as the work induced by quenching is $w=\pm Tr\rho
H_{\pm }\propto 1/\tau $.
\bibitem{diff} The absolute difference $\delta W$ and $\delta Q$ are
defined, respectively, as $\delta W=\left\vert W-W_{\mathrm{w}}\right\vert $
and $\delta Q=\left\vert Q-Q_{\mathrm{w}}\right\vert $, where the quantities
under weak SBC are $W_{\mathrm{w}}=\int\limits_{0}^{t}\mathrm{Tr}_{\mathrm{s}%
}(\rho _{\mathrm{s}}dH_{\mathrm{s}})$ and $Q_{\mathrm{w}}=\int\limits_{0}^{t}%
\mathrm{Tr}_{\mathrm{s}}(H_{\mathrm{s}}d\rho _{\mathrm{s}})$ with the state $%
\rho _{\mathrm{s}}$ chosen as the reduced state of the system coupled to its
bath.
\end{thebibliography}
\end{document}